\documentclass[twocolumn,secnumarabic,amssymb, nobibnotes, aps, prd]{revtex4-1}
\usepackage{graphicx}% Include figure files
\usepackage{amsbsy}
\usepackage{amsmath}

\begin{document}

\title{Linear response of light deformed nuclei investigated by self--consistent quasiparticle random--phase--approximation}%

\author{C. Losa$^{a}$, A. Pastore$^{b}$, T. D{\o}ssing$^{a}$, E. Vigezzi$^{d}$ and R. A. Broglia$^{a,c,d}$}
%\email[REVTeX Support: ]{revtex@aps.org}
\affiliation{
$^a$ The Niels Bohr Institute, University of Copenhagen, Blegdamsvej 17, 2100 Copenhagen \O, Denmark. \\
$^b$ Department of Physics, Post Office Box 35 (YFL), FI-40014 University of Jyv\"askyl\"a, Finland.\\
$^c$ Dipartimento di Fisica, Universit\`a degli Studi di Milano,via Celoria 16, 20133 Milano, Italy.\\
$^d$ INFN, Sezione di Milano, via Celoria 16, 20133 Milano, Italy.\\
}
\date{\today}%
\begin{abstract}
We present a calculation of the properties of vibrational states in deformed, axially--symmetric even--even nuclei,
within the framework of a fully self--consistent Quasparticle Random Phase  Approximation (QRPA). The same Skyrme
energy density and density-dependent pairing functionals are used to calculate the mean field and the residual
interaction in the particle-hole and particle-particle channels. We have tested our software in the case of spherical
nuclei against fully self consistent calculations published in the literature, finding excellent agreement. We
investigate the consequences of neglecting the spin-orbit and Coulomb residual interactions in QRPA. Furthermore we
discuss the improvement obtained in the QRPA result associated with the removal of spurious modes. Isoscalar and
isovector responses in the deformed ${}^{24}{}^{-}{}^{26}$Mg, ${}^{34}$Mg isotopes are presented and compared to
experimental findings.
\end{abstract}
\maketitle
%\tableofcontents

%%%% INTRODUZIONE %%%%%
\section{Introduction}\label{intro}

The response of many--body systems to an external, weakly coupled field provides much insight regarding
the correlations existing among the particles composing the system, and the forces acting among them.

While the most familiar application of mean field theory is to describe stationary states, its extension to
time--dependent states provides the basis for a theory of small amplitude oscillations known as Random Phase
Approximation (RPA) for normal systems and QRPA for superfluid (superconducting) systems displaying Quasiparticle
excitations.

Especially, small amplitude oscillations describe the dynamics of the nuclear surface. Couplings to the surface will
influence the quasiparticle motion, renormalizing the effective mass ($\omega$-mass $m_{\omega}$), in turn leading to
couplings among the different vibrational modes.

A textbook example of such couplings is provided by the breaking of the giant dipole resonance in deformed nuclei
(inhomogeneous damping \cite{Boh:75,Dat:87,Bor:98}), in keeping with the fact that a permanent deformation can be
viewed as a quadrupole vibration of finite inertia and of vanishing restoring force. Within this context we refer to
Fig. 10 of ref. \cite{carlos}, in particular to the two--peak photoabsorption cross section ($\sigma(E;E1)$) of the
$^{150}$Nd nucleus, marking the onset of static deformation in the Neodynium isotopes as a function of mass number.

As the smooth increase of the FWHM of $\sigma(E;E1)$ as a function of $A$ indicates, this cannot be a yes--or--no
effect. In fact, the FWHM of $\sigma(E;E1)$ associated with the transitional nucleus $^{148}$Nd is not very different
from that of $^{150}$Nd. Within this scenario, quadrupole deformations, static or dynamic, will also modify not only
the properties (centroid and width) of the GDR but also those of the GQR and, in deformed nuclei, that of the GMR. The
same line of reasoning also applies to other multipolarities of the static and dynamic deformations of the mean field,
which should not necessarily be only quadrupolar.

Because exotic nuclei, in particular neutron halo nuclei are, as a rule, more polarizable than nuclei lying along the
stability valley, one expects this to affect the modes and the associated renormalization to be especially important
for such nuclei. Here, one may mention the pygmy resonance \cite{pigmy} and the inversion of the usual sequence of
single particle energies \cite{Bar:01}. A consistent treatment of the vibrational modes is the first step on the way to
address such properties.

The low--frequency collective excitations are quite sensitive to the shell structure near the Fermi level as well as to
the nuclear surface shape, and one expects that new kinds of \mbox{collective} excitation will emerge under new
situations of nuclear structure. To investigate such possibilities, many calculations have been made using the
self-consistent RPA based on the Skyrme--Hartree--Fock (SHF) method \cite{Hama:96,Shlo:03} and the Quasiparticle--RPA
(QRPA) including pairing correlations \cite{Mats:01,Hagi:01,bender:02,Khan:02,Yama:04,Ter.ea:05}. A number of similar
approaches using different mean fields have also been carried out
\cite{Vret:01,Paar:03,Paar:04,Paar:05,Cao:05,Afa:05,Giambo:03,Sarchi:04}.
%Most of these calculations, however, are
%restricted to spherical nuclei.

Recently new iterative methods have been developed to calculate RPA strength functions for both spherical
\cite{doba:arx} and deformed \cite{Naka:07,Ina:09} nuclear systems. Low--frequency RPA modes in deformed nuclei close
to the neutron drip line have been studied \cite{Naka:05,Ina:06}, taking also into account pairing correlations
\cite{Urke:01,Alva:04}. These latter calculations are based on a BCS approximation which does not take into account
continuum coupling effects, typical of drip--line nuclei. A proper theoretical description of such weakly bound systems
requires a careful treatment of the asymptotic part of the nucleonic density. An appropriate framework for these
calculations is provided by the Hartree--Fock--Bogoliubov (HFB) formalism, solved in coordinate representation
\cite{doba:84,Doba.ea:05b} especially for spherical nuclei \cite{doba:95}, or more conveniently in the
configuration--space approach for deformed nuclei \cite{Doba.ea:05}.

A quantitative description of excitations in exotic nuclei is given by fully consistent QRPA calculations on top of an
HFB ground state, such that the same effective interaction is used for both calculations. A fully consistent HFB+QRPA
approach with the Gogny effective interaction for spherical and deformed nuclei has been developed in a harmonic
oscillator basis \cite{Peru.ea:08}. Standard QRPA equations have also been solved in a cylindrical box with the Skyrme
effective interaction, not including neither spin-orbit effects nor the Coulomb residual interactions
\cite{Yosh.ea:08,Yosh.Ya:08}.

%Recently, new iterative methods have been developed to calculate RPA strength functions \cite{Naka:07,Ina:09,doba:arx}.

Within this context we discuss in the present paper a consistent approach to describe linear response in deformed
nuclei within the framework of HFB+QRPA. Section \ref{metodo} discusses the elements used to work out a software to
implement such a program. In Sec. \ref{basic}, we provide detailed information regarding Hartree--Fock--Bogoliubov
ground states. Section \ref{basic} also illustrates some basic results of the QRPA software developed by us to treat
deformed nuclei when applied, for a consistency check, to the case of a spherical system. In this Section we also carry
out a discussion concerning the spurious modes. In Sec. \ref{str} the response functions of $^{20}$O, $^{24-26}$Mg and
$^{34}$Mg are shown and discussed in comparison to available data and to other calculations. Conclusions are drawn in
Sec. \ref{concludo}.

%%%%%%% METODO %%%%%%%%
\section{Method}\label{metodo}

As the first step in the self--consistent calculation of excitations in axially deformed and reflection symmetric
nuclei, HFB equations are solved in a finite harmonic oscillator (HO) as well as in a transformed harmonic oscillator
(THO) basis. In both cases, a discretization of the positive energy continuum is carried out. We use the new version
(known as (101)) of the program HFBTHO \cite{Doba.ea:05}, choosing for ${}^{20}$O, ${}^{24}{}^{-}{}^{26}$Mg and
${}^{34}$Mg, a quasiparticle energy cutoff $E_{cut}=50$ MeV, and set $N_{sh}=15$ HO (THO) shells. This code allows to
perform HFB calculations with arbitrary Skyrme functionals together with a density--dependent pairing delta interaction
\cite{Chas:76}

\begin{equation}\label{pairing int}
V_{pair} \left( {\textit{\textbf{r}}, \textit{\textbf{r}}'} \right)  = \frac{{1 - P_\sigma  }} {2}\left[ {V_0  +
\frac{{V_1 }} {6}\rho _{00}^\gamma  \left( \textit{\textbf{r}} \right)} \right]\delta \left( {\textit{\textbf{r}} -
\textit{\textbf{r}}'} \right),
\end{equation}

\noindent $\rho _{00}  \left( \textit{\textbf{r}}\right)$ being the associated isoscalar density and $P_\sigma$ the spin exchange
operator. In the following, we use the Skyrme functional SkM$^*$ \cite{Bartel:82} and a pairing interaction with the
same parametrization as that adopted in ref. \cite{Yosh.ea:08}, i.e. for ${}^{20}$O and ${}^{24}{}^{-}{}^{26}$Mg we adopt the
parameters $V_0=-280$ MeV fm$^3$, $V_1=-18.75 V_0$, $\gamma=1$ while for ${}^{34}$Mg $V_0=-295$ MeV fm$^3$, $V_1=-18.75
V_0$, $\gamma=1$ are employed. It corresponds to a mixed surface-volume type of pairing potential.

%The HFB program gives quasiparticle wave-functions in configuration space from which the single-particle density matrix
%$\rho$ can be constructed and subsequently diagonalized.

We diagonalize the HFB Hamiltonian in configuration space. We then diagonalize the density matrix $\rho$ obtaining the
canonical basis. We have checked the numerical accuracy of the procedure comparing the quasiparticle energies obtained
in the original diagonalization with those obtained diagonalizing the HFB Hamiltonian in the canonical basis. We have
found that the values agree within $10^{-4}$ MeV. Finally, we switch to coordinate space, tabulating the canonical
wave-functions together with their first and second derivatives in a grid of 30 $\times$ 30 (Gauss-Hermite) $\times$
(Gauss-Laguerre) points for Gauss quadrature integration in cylindrical coordinates
$(r_ \bot ,z)$. %$\textit{\textbf{r}} = \left( {r_ \bot ,z,\phi } \right)$.

The QRPA basis is obtained using pairs of these canonical wave functions such that the dimension of the
QRPA--Hamiltonian matrix does not exceed the size 20000 $\times$ 20000. For this purpose, one first omits the canonical
states $i$ that have single--particle energies $\varepsilon _i$ greater than some $\varepsilon _{crit}$. A second cut
is made excluding those QRPA quasiparticle pairs displaying occupation probabilities less than some small $v_{crit}^2$
or larger than $1 - v_{crit}^2$. In the following calculations these parameters are given the values $\varepsilon
_{crit}=200$ MeV and $v_{crit}=10^{-2}$. The QRPA excited states $\left| \lambda  \right\rangle$ are described in terms
of the quasi-boson operators

\begin{equation}\label{Qdaga}
Q_\lambda ^\dag   = \sum\limits_{K < K'} {\left( {X_{KK'}^\lambda  \alpha _K^\dag  \alpha _{K'}^\dag   -
Y_{KK'}^\lambda  \alpha _{K'} \alpha _K } \right)} ,
\end{equation}

\noindent acting on the QRPA correlated vacuum $|\tilde{0}\rangle \; (Q_\lambda |\tilde{0}\rangle = 0)$.

The operators ${\alpha _K^\dag  }$, ${\alpha _K }$ are the canonical quasiparticle creation and annihilation operators
respectively and ${X_{KK'}^\lambda  }$, ${Y_{KK'}^\lambda  }$ are the amplitudes of the two quasiparticles excitations
$\left\{ {K,K'} \right\}$. The matrix elements between different QRPA basis states $\left\{ {K,K'} \right\}$ and
$\left\{ {L,L'} \right\}$ are expanded respecting the selection rules of the vibration's quantum numbers $\Omega$ and
$\pi$. That is, $\Omega  = \Omega _K  + \Omega _{K'} = \Omega _L  + \Omega _{L'}$ and $\pi  = \pi _K \cdot \pi _{K'}  =
\pi _L \cdot \pi _{L'}$. Here, $\Omega$ is the projection of the angular momentum on the symmetry axis $z$ and $\pi$ is
the parity. Details concerning the calculation of matrix elements are given the Appendix.

%%%%%%% BASIC %%%%%%%%
\section{Basic results}\label{basic}

\def\thesubsection{\Alph{subsection}}
\subsection{HFB ground states}

First, potential energy curves are calculated as a function of the deformation parameter $\beta$, defined as

\begin{equation}\label{beta tot}
\beta  = \sqrt {\frac{\pi } {5}} \frac{{ < \hat Q > _n  +  < \hat Q > _p }} {{\left\langle {r^2 } \right\rangle _n  +
\left\langle {r^2 } \right\rangle _p }}.
\end{equation}

\noindent The quantity  $< \!\! \hat Q \!\! >_q$ is the average value of the quadrupole--moment operator $\hat Q = 2z^2
- r_ \bot ^2$ for protons ($q=p$) and neutrons ($q=n$). The QRPA calculation will be performed based on the HFB
solution corresponding to the absolute minimum of the potential energy.

\begin{figure}[!h]
\includegraphics[width=0.45\textwidth,angle=0]{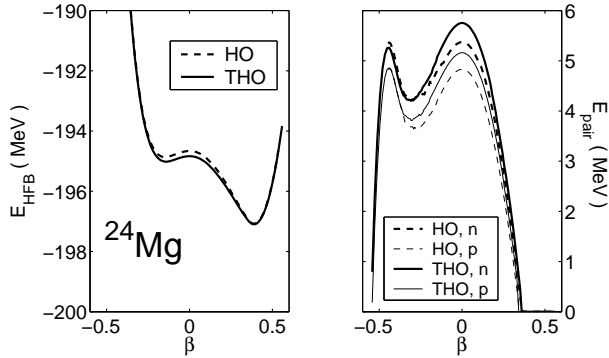}
\caption{HFB potential energy curves $E_{HFB}$ (left hand panel); neutron (thick lines) and proton (thin lines) pairing
energies $E_{pair}$ (right hand panel). The quantities are calculated  in HO (dashed curves) and THO (solid curves)
basis as functions of axial deformation parameter $\beta$ in the nucleus ${}^{24}$Mg.} \label{hfb mg24}
\end{figure}

Fig. \ref{hfb mg24} (left panel) shows the potential energy curves for the nucleus $^{24}$Mg, comparing the results
obtained in the HO and THO basis. One finds a pronounced minimum corresponding to the prolate deformation $\beta=
0.39$. Similar HFB calculations have been carried out for the other isotopes discussed in the following. The nucleus
$^{20}$O is found to be spherical, while $^{26}$Mg is oblate ($\beta= -0.18$) and $^{34}$Mg is prolate ($\beta= 0.36$).
The ground state properties of the four mentioned nuclei in the HO and THO basis are summarized in Table \ref{tab
HFB1}.

The right hand panel of Fig. \ref{hfb mg24} shows for ${}^{24}$Mg the pairing energy $E_{pair}  = \frac{1}
{2}\textrm{Tr}\left( {\Delta \kappa } \right)$ calculated separately for protons and neutrons, where $\kappa$ is the
expectation value ($\langle HFB | P^+ | HFB\rangle = \langle HFB |P| HFB \rangle$) in the superfluid ground state
($|HFB\rangle$) of the pair addition/removal ($P^+ = \sum_\nu c^+_\nu c^+_{\bar{\nu}} \;/\; P = \sum_\nu c_{\bar{\nu}}
c_\nu$, $c^+_\nu$: single--particle creation operator) operator, that is of the pair field (abnormal density). $\Delta$
stands for the functional derivative of the energy $E[\rho,\kappa]$ with respect to the abnormal density (pairing gap)
\cite{Doba.ea:05,Ring:80}.

Returning now to Fig. \ref{hfb mg24} one may notice the difference between the curves for protons and neutrons. The
Coulomb field influences the density of the nucleus and so the pairing interaction. At $\beta=0.39$ the total pairing
energy is zero, so QRPA calculations reduce to RPA ones where the excitations are only in the particle-hole (ph)
channel. Full QRPA calculations are performed in the other three nuclear systems ${}^{20}$O, ${}^{34}$Mg and
${}^{26}$Mg for which the HFB ground state displays neutron and proton pairing correlations respectively.

\begin{table*}
\begin{ruledtabular}
\begin{tabular}{lcccccccc}
                                                                  &\small{${}^{20}$O}   &\small{${}^{20}$O}   &\small{${}^{24}$Mg}  &\small{${}^{24}$Mg}  &\small{${}^{26}$Mg}   &\small{${}^{26}$Mg}   &\small{${}^{34}$Mg}  &\small{${}^{34}$Mg}   \\
                                                                  &\small{HO}           &\small{THO}          &\small{HO}           &\small{THO}          &\small{HO}            &\small{THO}           &\small{HO}           &\small{THO}           \\
\hline
\small{$\lambda_n$ (MeV)}                                         &-7.18                &-7.18                &-14.13               &-14.13               &-13.12                &-13.11                &-4.17                &-4.17                 \\
\small{$\lambda_p$ (MeV)}                                         &-17.27               &-17.25               &-9.51                &-9.51                &-11.05                &-11.03                &-20.19               &-20.17                \\
\small{$\beta_n \begin{array}{*{20}c} {} & {} & {}\end{array}$}   &0.0                  &0.0                  &0.38                 &0.38                 &-0.16                 &-0.16                 &0.37                 &0.37                  \\
\small{$\beta_p \begin{array}{*{20}c} {} & {} & {}\end{array}$}   &0.0                  &0.0                  &0.39                 &0.39                 &-0.16                 &-0.16                 &0.35                 &0.35                  \\
\small{$\left\langle \Delta  \right\rangle _n$ (MeV)}             &2.03                 &2.05                 &0.0                  &0.0                  &0.0                   &0.0                   &1.72                 &1.60                  \\
\small{$\left\langle \Delta  \right\rangle _p$ (MeV)}             &0.0                  &0.0                  &0.0                  &0.0                  &1.42                  &1.47                  &0.0                  &0.0                   \\
\small{$\sqrt {\left\langle {r^2 } \right\rangle _n}$ (fm)}       &2.91                 &2.91                 &2.99                 &3.00                 &3.01                  &3.01                  &3.50                 &3.51                  \\
\small{$\sqrt {\left\langle {r^2 } \right\rangle _p}$ (fm)}       &2.69                 &2.69                 &3.03                 &3.03                 &2.96                  &2.96                  &3.15                 &3.15                  \\
\small{$E_{HFB}$ (MeV)}                                           &-157.1               &-157.2               &-197.0               &-197.0               &-218.2                &-218.3                &-263.9               &-263.9                \\
\end{tabular}
\end{ruledtabular}
\caption{\footnotesize{Ground state properties of ${}^{20}$O, ${}^{24,26,34}$Mg obtained by the deformed HFB
calculation in the HO and THO basis. Chemical potentials $\lambda_q$, deformations $\beta_q$, average pairing gaps
$\Delta_q$, root--mean--square radii $\sqrt {\left\langle {r^2 } \right\rangle _q}$ for neutrons ($q=n$) and protons
($q=p$), and the total binding energies $E_{HFB}$ are listed.}}\label{tab HFB1}
\end{table*}

\subsection{Response for the spherical nucleus ${}^{48}$Ca}

To test our deformed QRPA code developed on the basis of the deformed HFBTHO program by M. Stoitsov \cite{Doba.ea:05},
we compare our results to those obtained with the spherical QRPA code developed by J. Terasaki $et$ $al.$ (JT)
\cite{Ter.ea:05}, making use of the same Skyrme interaction. Fig. \ref{fig str ca40tutto} shows, for the spherical
nuclear system ${}^{48}$Ca, such a comparison with calculations based on either {\it (i)} spherical basis states with a
hard-wall boundary condition at 20 fm, and {\it (ii)} the two types of deformed basis, HO and THO. The response
functions of Fig. \ref{fig str ca40tutto} have been calculated with the help of Eqs. (1), (2) of ref. \cite{Ter:06}. In
the following, we use strength functions defined as

\begin{equation}\label{lorenziana}
S_J^\tau  \left( E \right) = \sum\limits_\lambda  {\sum\limits_\Omega  {\frac{{\Gamma /2}} {\pi }} } \frac{{\left|
{\left\langle \lambda  \right|\hat F_{J\Omega }^\tau  \left| 0 \right\rangle } \right|^2 }} {{\left( {E - E_\lambda  }
\right)^2  + \Gamma ^2 /4}},
\end{equation}

\noindent for the multipole operator ${\hat F_{J\Omega }^\tau  }$. If not specified in the text, the value $\Gamma$ = 1
MeV is used as in ref. \cite{Yosh.ea:08} in the calculation of the IS monopole and quadrupole transition operators

\begin{equation}\label{F quadrupolo IS}
\hat F_{2\Omega }^{IS}  = \frac{{eZ}} {A}\sum\limits_{i = 1}^A {r_i^2 Y_{2\Omega } \left( {\hat r_i } \right),}
\end{equation}

\begin{equation}\label{F monopolo IS}
\hat F_{00}^{IS}  = \frac{{eZ}} {A}\sum\limits_{i = 1}^A {r_i^2 ,}
\end{equation}

\noindent and for the IV dipole operator

\begin{equation}\label{F dipolo IV}
\hat F_{1\Omega }^{IV}  = \frac{{eN}} {A}\sum\limits_{i = 1}^Z {r_i Y_{1\Omega } \left( {\hat r_i } \right)}  -
\frac{{eZ}} {A}\sum\limits_{i = 1}^N {r_i Y_{1\Omega } \left( {\hat r_i } \right)} .
\end{equation}

\noindent The agreement between the strength functions labeled JT \cite{Ter:06} with those labeled HO and THO shown on
Fig. \ref{fig str ca40tutto} is actually quite rewarding, in keeping with the completely different routes taken in the
calculations. In all the three cases the strength function exhausts about 98 $\%$ of the Energy Weighted Sum Rule
(EWSR).

For the 2$^+$ mode in ${}^{48}$Ca, the degeneracy between the components $\Omega^\pi=0^+, \pm 1^+, \pm 2^+$ is reached
up to the order of 10$^{-3}$ in both the HO and THO basis.

%It has been essential for the development of the deformed code to have the spherical code to compare to, term by term
%of the residual interaction.

%All together, we could notice the overall good agreement between both the HO and THO approaches with the cylindrical
%coordinates and that in spherical coordinates of JT who uses a completely different basis and calculation scheme
%\cite{Losa:2009}. To try to understand the relatively small discrepancies among the three calculations, the next step
%could be to inspect the calculated peaks of the giant resonance more carefully, eventually by comparing the main
%quasiparticle pairs of excitations, and by comparing the distribution of transition strength over the nuclear volume.

%\begin{figure}[!!h]
%\begin{center}
%\includegraphics*[width=0.38\textwidth,angle=0]{figure/wf_ca48}
%\end{center}
%\caption{\footnotesize{Spin--up component of selected $d_{3/2}$ states with the $\Omega ^\pi   =  \frac{1} {2}^ +$
%projection in ${}^{48}$Ca. Wave functions are calculated in the spherical basis by the code HFBRAD for $\theta$ = 0
%(dashed--dotted curve) at canonical single--particle energy $\varepsilon=11.08$ MeV and in the HO basis by the present
%program fixing $r_\perp$ = 0 (dotted curve) and $z$ = 0 (solid curve) at $\varepsilon=13.23$ MeV.}} \label{fig d3/2ho
%ca40}
%\end{figure}

\begin{figure}[!h]
\begin{center}
\includegraphics*[width=0.38\textwidth,angle=0]{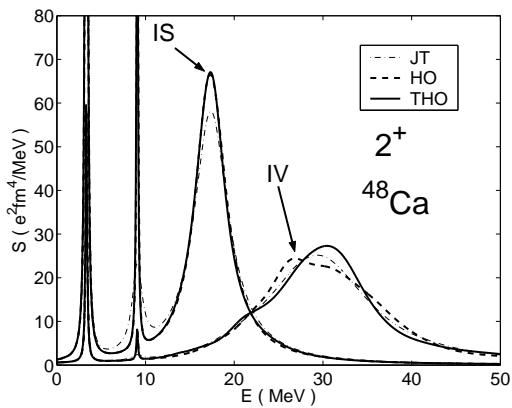}
\end{center}
\caption{\footnotesize{Isoscalar (IS) and isovector (IV) $2^+$ strength functions in ${}^{48}$Ca in the spherical basis
of J. Terasaki (JT, dashed--dotted curve), in HO (dashed curve) and THO (solid curve) basis.}} \label{fig str
ca40tutto}
\end{figure}

\subsection{Self-consistency and spurious states}

Let us now turn to the spurious solutions of the QRPA equations in the dipole modes with $\Omega^\pi=0^-, \pm 1^-$ and
in the quadrupole modes with $\Omega^\pi=\pm 1^+$. Spurious modes are, in the present case, due to the translational
and rotational symmetry breaking, respectively, of the HFB ground state \cite{Ring:80}. The quadrupole modes with
$\Omega^\pi=0^+$ also contains spurious states associated with particle--number nonconservation. Former works on QRPA
both for spherical nuclei \cite{Ter.ea:05} and for deformed ones \cite{Peru.ea:08} show that the spurious states which
should be at zero energy \cite{Thoul:60} have instead values from a few keV to 1--1.8 MeV, well separated from, in any
case, (physical) states. These nonzero energies may be taken as a measure of numerical accuracy with which linear
response is calculated, and are strongly related to the choice of basis states and truncation of basis size.

\begin{figure}[!h]
\begin{center}
\includegraphics*[width=0.38\textwidth,angle=0]{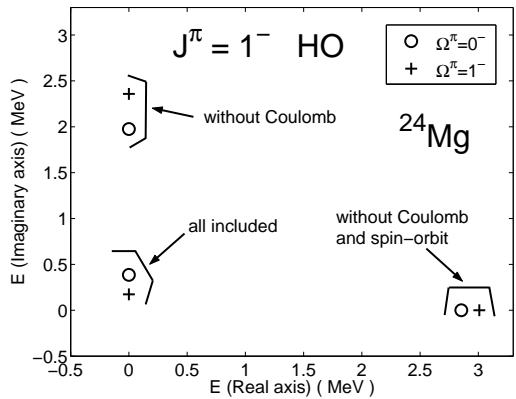}
\end{center}
\caption{\footnotesize{Energies on the complex plane of the spurious state for the $1^-$ mode with $\Omega^\pi$=0$^-$
(circles) and $\Omega^\pi$=1$^-$ (crosses) projections in ${}^{24}$Mg in HO basis. The figure illustrates the change in
energy of spurious modes when specific terms of the interaction are omitted.}} \label{spurio1}
\end{figure}

\begin{figure}[!h]
\begin{center}
\includegraphics*[width=0.38\textwidth,angle=0]{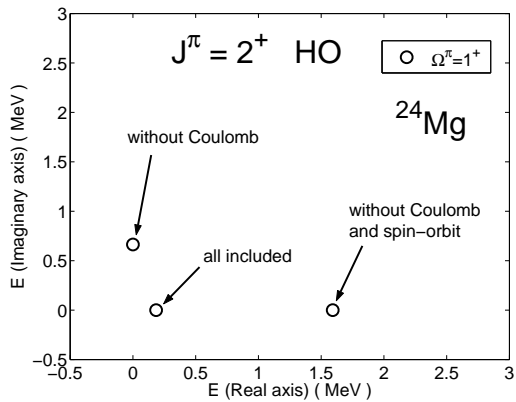}
\end{center}
\caption{\footnotesize{The same as Fig. \ref{spurio1} but for the $2^+$ mode with $\Omega^\pi$=1$^+$ projection.}}
\label{spurio2}
\end{figure}

The spurious states obtained in our fully consistent calculations lie either on the real on the imaginary  axis, and
their absolute value never exceeds 700 keV. In Fig. \ref{spurio1} and \ref{spurio2} we show results for $^{24}$Mg,
comparing the self-consistent results with those obtained (i) neglecting both Coulomb and spin--orbit residual
interaction and (ii) neglecting only the Coulomb residual interaction. It is seen that the absolute values of the
energies are progressively reduced going from case (i) (about 3 MeV) to case (ii) (2-2.2 MeV) and finally to the
self-consistent solution (less than 0.5 MeV). One also notes that in case (i) the energies of the spurious states lie
on the real axis, while in case (ii) they are found on the imaginary axis. The fully consistent QRPA calculations give
imaginary energies for the dipole spurious modes and real energies for the quadrupole spurious modes, all below 500
keV. It is satisfactory that the energies of the spurious modes get closer to zero as more terms of the residual
interaction are included. We ascribe the remaining distance to zero to the truncation of the basis and to numerical
inaccuracies. Within this scenario one may posit that a spurious mode obtained by fully consistent QRPA calculations at
"imaginary energy" 100--500 keV is as good a solution as that corresponding to a spurious real energy mode at 100--500
keV.

A similar study has been performed on the energies of the dipole spurious modes of ${}^{20}$O. Also for this nucleus if
both Coulomb and spin--orbit parts are neglected, the spurious energies are at 2.5--3 MeV. The values jump to about
1.5--2 MeV on the imaginary axis if one adds the spin--orbit contribution to the residual interaction. Finally, the
energy of a spurious state obtained with a fully consistent QRPA calculation can be real or imaginary not exceeding 600
keV. For the deformed nuclei ${}^{26}$Mg and ${}^{34}$Mg we obtain from the fully consistent QRPA calculations with
$\Omega^\pi=0^-, \pm 1^-$ and $\Omega^\pi=\pm 1^+$ spurious states with either real or imaginary energy, in any case
the modulus not exceeding 700 keV. For ${}^{20}$O and ${}^{26,34}$Mg the quadrupole spurious modes with
$\Omega^\pi=0^+$ are all real at around 1.8--2 MeV. We ascribe this problem concerning the $\Omega^\pi=0^+$ modes
mainly to the rather crude cut in the occupation probability given by $v_{crit}=10^{-2}$. We have checked in the case
of $^{26}$Mg that reducing the value of $v_{crit}$ to $10^{-3}$ brings the energy of the $0^+$ spurious state from 1.8
MeV (on the real axis) down to 1.1 MeV (on the imaginary axis). The transition strength is only affected for energies
below about 3 MeV.

Similar investigations of the spurious modes for calculations with the Gogny force have been reported by S. Peru $et$
$al.$ in references \cite{Peru.ea:pf} and \cite{Peru.ea:08} for spherical and deformed shapes, respectively. For
spherical nuclei, it is found that the spurious modes lie at very low energy, about 3-5 keV, when all terms of the
interaction are included, whereas they may move up to about 2 MeV when leaving out parts of the interaction. For
deformed nuclei \cite{Peru.ea:08}, the result is qualitatively the same as in the present work. However, the Gogny
results are better for the $\Omega^{\pi} = 0^{+}$ mode, which comes at a very low energy, but worse for the
translational invariance modes $\Omega^{\pi} = 0^{-},\pm 1^{-}$, where the spurious mode energies can come as high as
1.8 MeV.

%The solutions of the QRPA equations are used to calculate the strength functions

%\begin{equation}\label{lorenziana}
%S_J^\tau  \left( E \right) = \sum\limits_\lambda  {\sum\limits_\Omega  {\frac{{\Gamma /2}} {\pi }} } \frac{{\left|
%{\left\langle \lambda  \right|\hat F_{J\Omega }^\tau  \left| 0 \right\rangle } \right|^2 }} {{\left( {E - E_\lambda  }
%\right)^2  + \Gamma ^2 /4}},
%\end{equation}

%\noindent for the multipole operator ${\hat F_{J\Omega }^\tau  }$.

%If not specified in the text, in the following calculations we use $N_0$ = 15 HO and THO major shells, the Skyrme
%functional SKM$^*$, a quasiparticle--energy cutoff $E_{cut}$ = 60 MeV, a single particle energy cutoff $\varepsilon
%_{crit}$ = 200 MeV and a quasiparticle occupation cutoff $v_{crit}^2$ = $10^{ - 4}$.

%%%%% RISULTATI %%%%%%
\section{Strength functions}\label{str}

\def\thesubsection{\Alph{subsection}}
\subsection{${}^{20}$O}

\begin{figure}[!h]
\begin{center}
\includegraphics*[width=0.38\textwidth,angle=0]{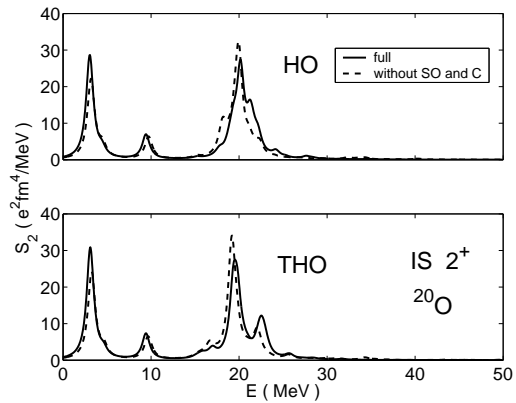}
\end{center}
\caption{\footnotesize{IS $2^+$ strength functions in ${}^{20}$O, in HO (upper panel) and THO (lower panel) basis
without the residual spin-orbit (SO) and Coulomb (C) interaction (dashed curve) and with all the terms included (solid
curve).}} \label{fig O20is2J}
\end{figure}

The IS quadrupole $2^+$ mode is shown in Fig. \ref{fig O20is2J}, calculated with the HO as well as with the THO basis
states. A large fraction of the strength function is concentrated in three energy regions: a low-lying mode around 3
MeV, a small peak around 10 MeV, and a giant resonance around 20 MeV. The two calculations display a remarkable
agreement with respect to the strength function below 15 MeV. Whereas the calculations agree on the centroid and the
strength of the giant resonance, they display differences with respect to the splitting of the two components of the
resonance, and on the distribution of strength between the two peaks. When we omit both the spin-orbit and Coulomb
residual interaction, in the giant resonance region the strength functions are shifted down in energy about 0.3 MeV for
both the basis used, and the peak heights increase by 15$\%$ for HO and 25$\%$ for THO. Concerning the two peak
energies below 10 MeV, they are both shifted up by about 0.2 MeV while the peak heights are lowered by 20$\%$ for both
the two basis.

\begin{figure}[!h]
\begin{center}
\includegraphics*[width=0.38\textwidth,angle=0]{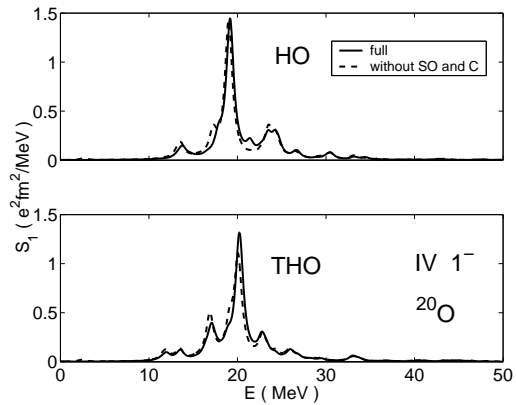}
\end{center}
\caption{\footnotesize{The same as Fig. \ref{fig O20is2J} but for IV $1^-$ strength functions.}} \label{fig O20iv1J}
\end{figure}

Fig. \ref{fig O20iv1J} shows the IV giant dipole resonance in the energy range (15-25) MeV. In the THO basis the peak
is at 20.2 MeV, while that in the HO basis is shifted down about 1 MeV.

The strength functions calculated without the spin-orbit and Coulomb interactions  can be compared to those obtained by
K. Yoshida $et$ $al.$ \cite{Yosh.ea:08}, who did not include these terms. We find a very good overall agreement. On
closer inspection, the energy of the giant resonance peaks in our calculation in the THO basis is about 0.3 MeV higher
than in ref. \cite{Yosh.ea:08}.

\subsection{${}^{24,26}$Mg}

\subsubsection{IS 2$^+$ modes}

\begin{figure}[!!h]
\begin{center}
\includegraphics*[width=0.38\textwidth,angle=0]{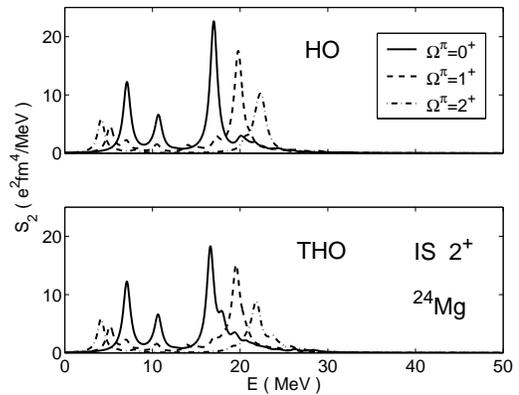}%cut 50
\end{center}
\caption{\footnotesize{IS $2^+$ strength functions in ${}^{24}$Mg, in HO (upper panel), and THO (lower panel) basis for
the $\Omega^\pi$=$0^+$ (solid curve), $1^+$ (dashed curve) and $2^+$ (dashed-dotted curve) excitations.
%The calculated sum satisfies 95.84 $\%$ and
%95.88 $\%$ of the EWSR value in the HO and THO basis respectively.
}} \label{fig Mg24 2ISph}
\end{figure}

In Figs. \ref{fig Mg24 2ISph}, \ref{fig Mg26 2IS} we display the response functions of the IS quadrupole mode of the
deformed prolate and oblate ${}^{24,26}$Mg isotopes.

For both these nuclei, the first low lying state is found to have projection $\Omega=2^{+}$, corresponding to a
$\gamma$-vibration, while the $\beta$-vibration with $\Omega=0^{+}$ is situated at a considerably higher energy.
Comparing the nuclei, one sees that $^{24}$Mg, which may be described as a deformed closed shell nucleus for the
prolate shape, is more stiff towards vibrations than the oblate $^{26}$Mg. This is all in qualitative accordance with
data \cite{Endt}. However, a more careful comparison of calculations and data reveal some discrepancies, especially for
$^{26}$Mg.

In $^{24}$Mg, the calculated energy of the lowest $\Omega^{\pi} = 2^{+}$ is 4.14 MeV, to be compared to the
experimental value 4.23 MeV of the gamma--band head, the state $2^{+}_2$ \cite{Endt}. Likewise, the calculated energy
of the lowest $\Omega^{\pi} = 0^{+}$ is 7.08 MeV, should be compared to the experimental value 6.43 MeV. Both these
states are not very collective, composed equally of proton and neutron particle-hole excitations: $\left[ {N, n_z ,
\Lambda , \pm \Omega ^\pi } \right] \equiv \left[ {2,1,1, \pm \frac{3} {2}^ + } \right] \to \left[ {2,1,1, \pm \frac{1}
{2}^ + } \right]$ for the $\Omega = 2^+$ state and $\left[ {2,2,0, \pm \frac{1} {2}^ + } \right] \to \left[ {2,1,1, \pm
\frac{1} {2}^ + } \right]$ for the $\Omega = 0^{+}$ state, respectively. Still, the RPA root is shifted down in energy
by about 1.5 MeV relative to the single ph excitations. These assignments are in reasonable accordance with
experimental information from $\gamma-$decay \cite{Endt}, particle transfer \cite{Garrett}, (e,e') inelastic scattering
\cite{Zarek}, ($\alpha$,$\alpha'$) scattering \cite{nilson} and ($\pi$,$\pi'$) scattering~\cite{Blanpied}.

In $^{26}$Mg, the lowest $\Omega^{\pi} = 2^{+}$ excitation is calculated to be at 1.31 MeV, considerably below the
experimental state $2^{+}_2$ at 3.0 MeV, while the lowest $\Omega^{\pi} = 0^{+}$ excitation is at 2.75 MeV, to be
compared to the state $0^{+}_2$ at 3.64 MeV. Compared to $^{24}$Mg, the $\Omega^{\pi} = 2^{+}$ excitation is predicted
to be more collective. This is not in accordance with the experimental B(E2) values from the ground state, which are of
about equal magnitude for the two nuclei. Also, the considerable matrix element for exciting the $\Omega^{\pi} = 0^{+}$
state predicted by the calculations is in disagreement with the long life--time seen experimentally of the $0^{+}_2$
state.

The rather low collectivity of the $\beta$- and $\gamma$-vibrations follows a general trend. With a permanent
deformation, most of the quadrupole collectivity is tied up in the deformation of the mean field. This will in turn
influence the renormalization effects these modes will have on the single particle motion. In this context, one can
quote the rather different pattern observed in the distribution of matrix elements contributing to the induced pairing
interaction in spherical and deformed nuclei, respectively \cite{Bar:99,Don:xx}.

Next, we turn our attention to the giant vibrations situated in the energy region 15-25 MeV for all cases, with a
characteristic splitting between the different projections $\Omega$. For the prolate nucleus ${}^{24}$Mg, the
$\Omega^\pi = 0^+$ vibration is along the longest axis and acquires the lowest energy. For the oblate nucleus
${}^{26}$Mg, the opposite behavior is observed, and here the splitting is less pronounced due to the smaller value of
the deformation parameter \cite{Bor:98}.

In Tables \ref{tab1mg24}, \ref{tab1mg26} the mean energies

\begin{equation}\label{mean energy}
E(GR) = \frac{{m_1 \left[ {E_{\min } ,E_{\max } } \right]}} {{m_0 \left[ {E_{\min } ,E_{\max } } \right]}},
\end{equation}

\noindent with moments

\begin{equation}\label{moments mean energy}
m_\lambda  \left[ {E_{\min } ,E_{\max } } \right] = \sum\limits_k {\sum\limits_\Omega  {E_k^\lambda  \left|
{\left\langle k \right|\hat F_{J\Omega } \left| 0 \right\rangle } \right|^2 } } ,
\end{equation}

\noindent calculated for the 2$^+$ modes of ${}^{24,26}$Mg in the energy range ${\left[ {E_{\min } ,E_{\max } }
\right]}$ = ${\left[9,41\right]}$ MeV, are given in HO and THO basis. First, one sees that the mean value in the giant
resonance region is practically not affected by the choice of basis.  Secondly, one may address the 1 MeV difference
between the results obtained with the SkM$^*$ Skyrme force and those with the Gogny force, by S. P\'{e}ru $et$ $al.$
\cite{Peru.ea:08}. Part of this difference may be caused by the lower effective mass $m^* /m$ = 0.7 for the Gogny
force, as compared to $m^* /m$ = 0.8 for the SkM$^*$ force.  For this reason we performed for ${}^{24}$Mg a calculation
also with a SLy4 force where $m^*/m$ = 0.7. The centroid energy then increases by roughly 0.6 MeV, indicating that
about half of the difference between the two calculations can be ascribed to the effective mass. Finally, the results
may be compared to the experimental mean energy of 16.9 $\pm$ 0.6 obtained by Youngblood $et$ $al.$ \cite{Lui.ea:99},
and one sees that all the theoretical calculations overshoot this value by at least 2 MeV.

\begin{table}
\begin{ruledtabular}
\begin{tabular}{lccc}
\small{${}^{24}$Mg}                       &\small{IS}                  &\small{IS}                     &\small{IV}                     \\
\small{$E(GR)$}                           &\small{J$^\pi=0^+$}         &\small{J$^\pi=2^+$}            &\small{J$^\pi=1^-$}            \\
\hline
\small{Interval}                          &\small{[9,41]}              &\small{[9,41]}                 &\small{[10,29]}                \\
\hline
\small{Theor. D1S (P\'{e}ru $et$ $al.$)}  &21.0                        &20.5                           &23.0                           \\
\small{Theor. SkM$^*$ (HO)}               &20.7                        &19.4                           &19.6                           \\
\small{Theor. SkM$^*$ (THO)}              &20.3                        &19.3                           &19.8                           \\
\small{Theor. SLy4 (HO)}                  &                            &20.0                           &19.8                           \\
\small{Theor. SLy4 (THO)}                 &                            &19.9                           &19.8                           \\
\small{Exp. (Youngblood $et$ $al.$)}      &21.0 $\pm$ 0.6              &16.9 $\pm$ 0.6                 &                               \\
\small{Exp. (Irgashev $et$ $al.$)}        &                            &                               &22.1                           \\
\end{tabular}
\end{ruledtabular}
\caption{\label{tab1mg24} Theoretical mean energy values (in MeV) obtained with the Gogny force by S. P\'{e}ru $et$
$al.$, with the SkM$^*$ and SLy4 Skyrme force in HO and THO basis (present work), and experimental mean energy values
by Youngblood $et$ $al.$ and Irgashev $et$ $al.$ of the IS giant monopole, quadrupole resonances (calculated in the
energy interval ${\left[ {E_{\min } ,E_{\max } } \right]}$ = ${\left[9,41\right]}$ MeV) and IV giant dipole resonances
(calculated in the energy interval ${\left[ {E_{\min } ,E_{\max } } \right]}$ = ${\left[10,29\right]}$ MeV) in
${}^{24}$Mg.}
\end{table}

\begin{table}
\begin{ruledtabular}
\begin{tabular}{lccc}
\small{${}^{26}$Mg}                       &\small{IS}                   &\small{IS}                     &\small{IV}                     \\
\small{$E(GR)$}                           &\small{J$^\pi=0^+$}          &\small{J$^\pi=2^+$}            &\small{J$^\pi=1^-$}            \\
\hline
\small{Interval}                          &\small{[9,41]}               &\small{[9,41]}                 &\small{[10,29]}                \\
\hline
\small{Theor. D1S (P\'{e}ru $et$ $al.$)}  &22.0                         &21.0                           &22.9                           \\
\small{Theor. SkM$^*$ (HO)}               &22.1                         &20.3                           &20.3                           \\
\small{Theor. SkM$^*$ (THO)}              &21.9                         &20.3                           &20.2                           \\
\small{Exp. (Fultz $et$ $al.$)}           &                             &                               &20.6                           \\
\end{tabular}
\end{ruledtabular}
\caption{\label{tab1mg26} Theoretical mean energy values (in MeV) obtained with the Gogny force by S. P\'{e}ru $et$
$al.$, with the SkM$^*$ Skyrme force in HO and THO basis (present work) of the IS giant monopole, quadrupole resonances
(calculated in the energy interval ${\left[ {E_{\min } ,E_{\max } } \right]}$ = ${\left[9,41\right]}$ MeV), and of the
IV giant dipole resonance (calculated in the energy interval ${\left[ {E_{\min } ,E_{\max } } \right]}$ =
${\left[10,29\right]}$ MeV) in ${}^{26}$Mg. For this last resonance the experimental mean energy value by Fultz $et$
$al.$ is also given.}
\end{table}

The theoretical results obtained in the present work with the SkM$^*$ and SLy4 forces exhaust 83--84 $\%$ of the EWSR
exceeding the experimental result by about 20~$\%$. This difference in the EWSR as well as in the mean energy between
the present work and the experimental data of Youngblood $et$ $al.$, is apparent from Fig. \ref{fig Mg24 exp2}, which
shows the comparison between calculated and experimental strength functions. The shape of the experimental curve is
reproduced in its overall features, in particular for $\Gamma$=3 MeV. However, according to the calculations, the
central peak around 20 MeV is too pronounced, and there is too little strength towards lower energies. It is of notice
the sharp peak experimentally observed at 15 MeV. To which extent it may be connected with the $\Omega^\pi = 0^+$ mode
which in our calculations appear blue shifted by 2.5 MeV (i.e. at 17.5 MeV) is an open question.

Fig. \ref{fig Mg24 no soC} shows for ${}^{24}$Mg the effect of leaving out the spin--orbit and Coulomb parts of the
residual interaction. One sees that this would lead to a downward shift of the giant resonance by about 0.9 MeV.
Comparing to the work of K. Yoshida $et$ $al.$ \cite{Yosh.ea:08} one finds that their peaks are shifted further down by
about 0.9 MeV with respect to those obtained in the present work without the spin--orbit and Coulomb terms. This
remaining shift should be compared to the equivalent shift of 0.3 MeV discussed above for ${}^{20}$O. It should be
ascribed to the differences between the two calculations, the renormalization of the interaction in ref.
\cite{Yosh.ea:08}, and the different basis used.

\begin{figure}[!!h]
\begin{center}
\includegraphics*[width=0.38\textwidth,angle=0]{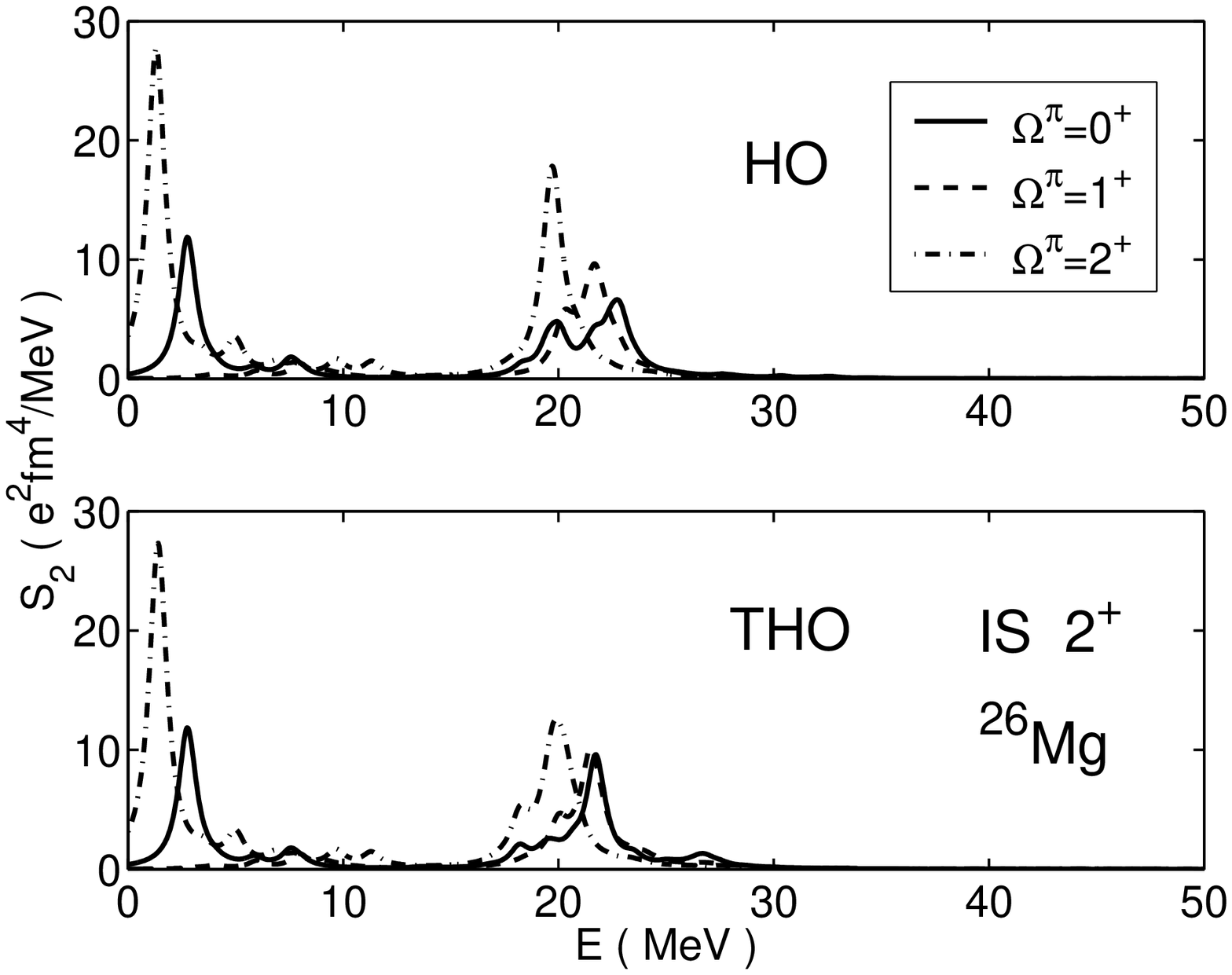}%cut 50
\end{center}
\caption{\footnotesize{IS $2^+$ strength functions in ${}^{26}$Mg, in HO (upper panel), and THO (lower panel) basis for
the $\Omega^\pi$=$0^+$ (solid curve), $1^+$ (dashed curve) and $2^+$ (dashed-dotted curve) excitations.}} \label{fig
Mg26 2IS}
\end{figure}

\begin{figure}[!!h]
\begin{center}
\includegraphics*[width=0.38\textwidth,angle=0]{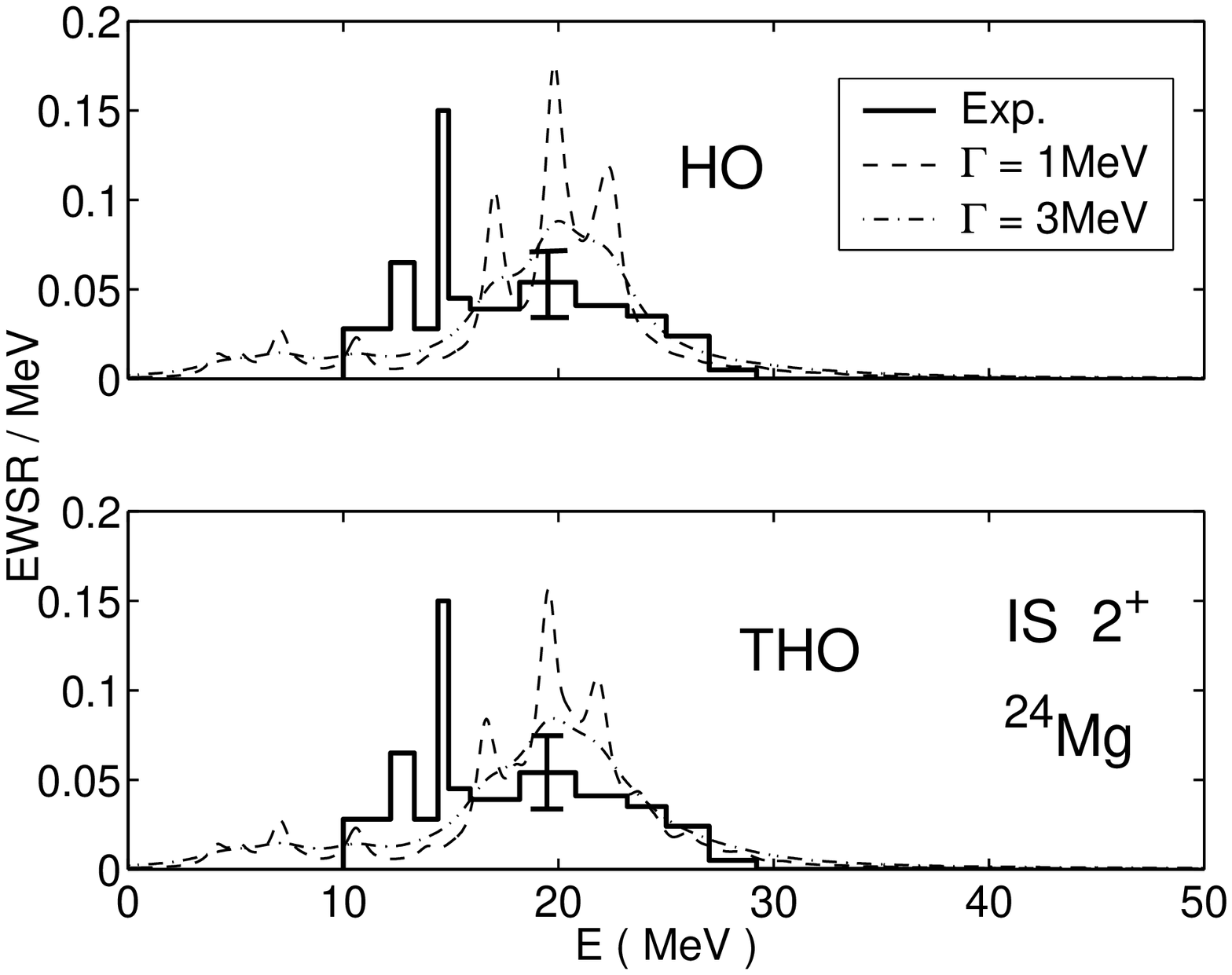}%cut 50
\end{center}
\caption{\footnotesize{Fractions of IS quadrupole EWSR in ${}^{24}$Mg, in HO (upper panel) and THO (lower panel) basis.
Dashed and dot-dashed curves are obtained from folding of QRPA spectra with a Lorentian distribution having $\Gamma$=1
MeV and $\Gamma$=3 MeV respectively. These are compared with the experimental data by Youngblood $et$ $al.$ (solid
curve).}} \label{fig Mg24 exp2}
\end{figure}

\begin{figure}[!!h]
\begin{center}
\includegraphics*[width=0.38\textwidth,angle=0]{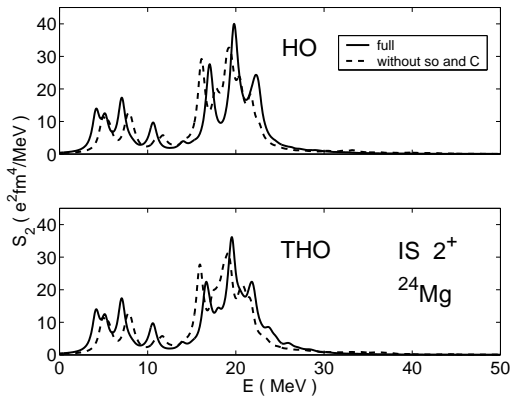}%cut 50
\end{center}
\caption{\footnotesize{IS $2^+$ strength functions in ${}^{24}$Mg, in HO (upper panel) and THO (lower panel) basis
without the residual spin-orbit (SO) and Coulomb (C) interaction (dashed curve) and with all the terms included (solid
curve).}} \label{fig Mg24 no soC}
\end{figure}

\subsubsection{IV 1$^-$ modes}

\begin{figure}[!!h]
\begin{center}
\includegraphics*[width=0.38\textwidth,angle=0]{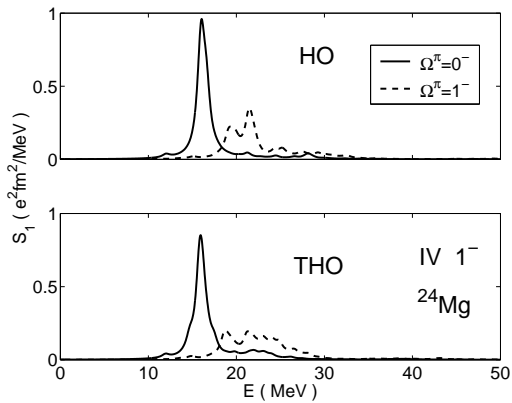}%cut 50
\end{center}
\caption{\footnotesize{IV $1^-$ strength functions in ${}^{24}$Mg, in HO (upper panel), and in THO (lower panel) basis
for the $\Omega^\pi$=$0^-$ (solid curve), $1^-$ (dashed curve) excitations.
%The calculated sum satisfies 96.870 $\%$ and 96.873 $\%$ of the EWSR value in the HO and THO basis respectively.
}} \label{fig Mg24 1IVph}
\end{figure}

\begin{figure}[!!h]
\begin{center}
\includegraphics*[width=0.38\textwidth,angle=0]{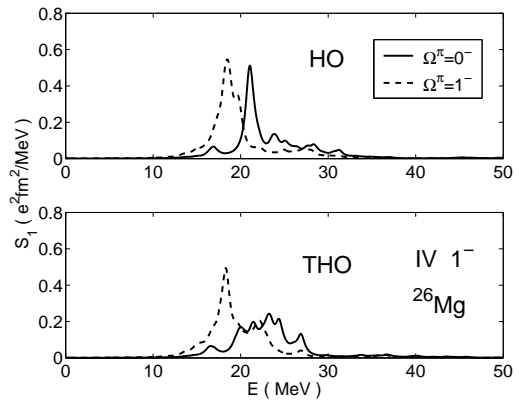}%cut 50
\end{center}
\caption{\footnotesize{IV $1^-$ strength functions in ${}^{26}$Mg, in HO (upper panel), and in THO (lower panel) basis
for the $\Omega^\pi$=$0^-$ (solid curve), $1^-$ (dashed curve) excitations.}} \label{fig Mg26 1IV}
\end{figure}

Figs. \ref{fig Mg24 1IVph} and \ref{fig Mg26 1IV} show the response functions of the IV dipole modes, of the deformed
prolate and oblate ${}^{24,26}$Mg isotopes. The IV giant dipole resonances show a two--peaked structure. For both
nuclear systems the low--lying part of the resonances is given by a defined peak at around 16 MeV and 18 MeV for
${}^{24}$Mg and ${}^{26}$Mg respectively. The higher energy part of the strength is fragmented, especially for the THO
basis, in fair agreement with the responses given by S. P\'{e}ru $et$ $al.$ \cite{Peru.ea:08}. In the HO approach one
may see a two--peaked structure up to around 26 MeV for ${}^{24}$Mg and a defined peak at 21--22 MeV for ${}^{26}$Mg.

\begin{figure}[!!h]
\begin{center}
\includegraphics*[width=0.38\textwidth,angle=0]{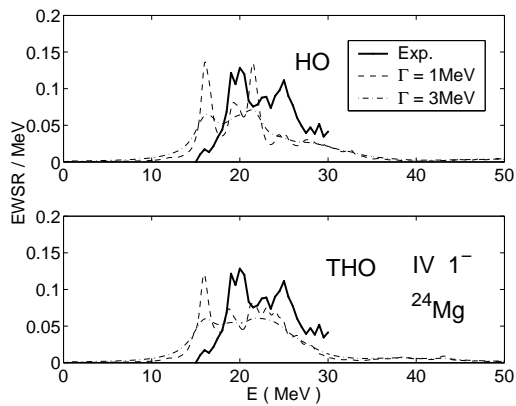}%cut 50
\end{center}
\caption{\footnotesize{Fractions of IV dipole EWSR in ${}^{24}$Mg, in HO (upper panel), and in THO (lower panel) basis.
Dashed and dot-dashed curves are obtained from folding of QRPA spectra with a Lorentian distribution having $\Gamma$=1
MeV and $\Gamma$=3 MeV respectively. These are compared with the experimental data by Irgashev $et$ $al.$ (solid
curve).}} \label{fig Mg24 exp1iv}
\end{figure}

The fraction of the EWSR in ${}^{24}$Mg for the IV 1$^-$ mode calculated in the present work for two values of $\Gamma$
is compared to the experimental ones \cite{Ish:88} in Fig. \ref{fig Mg24 exp1iv}. The theoretical and experimental
curves show the same two--peak structure, but the calculated peaks appear at energies which are about 3 MeV too low.
This difference with the experimental dipole response is also found in the calculations by T. Inakura $et$ $al.$
\cite{Ina:09}. In fact, the experimental mean energy that we extract from the data in the range ${\left[ {E_{\min }
,E_{\max } } \right]}$ = ${\left[15,30\right]}$ MeV is equal to 22.1 MeV, to be compared with our value of 19.8 MeV.
The Gogny--force calculation by S. P\'{e}ru $et$ $al.$ \cite{Peru.ea:08} instead displays the peaks at about the
observed energies, with a mean energy of 23.0 MeV (cf. Table \ref{tab1mg24}).

One may comment that the excitation energies of the lowest peak predicted by our calculation (17 MeV) is just above the
threshold for particle emission. At these energies, a considerable part of photon absorption events will lead to a
$\gamma$-ray cascade rather than particle emission. In fact, the ${}^{24}$Mg$\left( {e ,e'} \right)$ inelastic
scattering experiment by Titze $et$ $al.$ \cite{Titze:67} displays a peak around 17 MeV (split into two rather narrow
components), whereas the $\left( {\gamma ,n} \right)$ and $\left( {\gamma ,p} \right)$ data only display a tiny and
statistically insignificant peak at this energy. On the other hand, $\left( {e ,e'} \right)$ reactions are more
complicated to analyze theoretically in a precise way than hadronic inelastic processes and photon absorption, since a
consistent calculation of the cross section should go beyond the long wavelength approximation, and also take into
account magnetic interactions.

\begin{figure}[!!h]
\begin{center}
\includegraphics*[width=0.38\textwidth,angle=0]{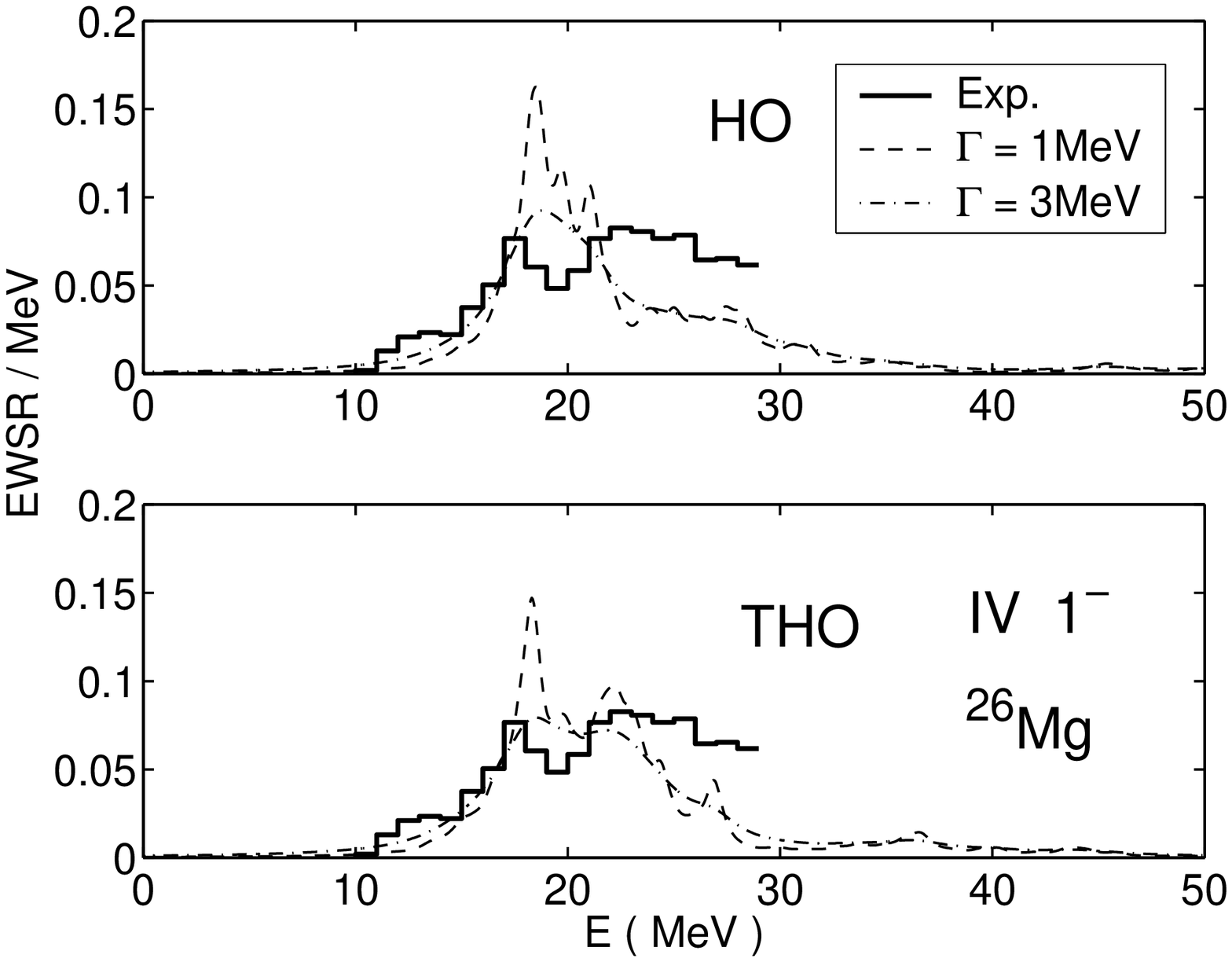}%cut 50
\end{center}
\caption{\footnotesize{Fractions of IV dipole EWSR in ${}^{26}$Mg, in HO (upper panel), and in THO (lower panel) basis.
Dashed and dot-dashed curves are obtained from folding of QRPA spectra with a Lorentian distribution having $\Gamma$=1
MeV and $\Gamma$=3 MeV respectively. These are compared with the experimental data by Fultz $et$ $al.$ (solid curve).}}
\label{fig Mg26 1IV fraz}
\end{figure}

For ${}^{26}$Mg, the fractions of the EWSR may be compared to three experiments \cite{Titze:70,Fultz:71,Ish:72}. Fig.
\ref{fig Mg26 1IV fraz} displays the comparison between our curves and that of Fultz $et$ $al.$ \cite{Fultz:71} who,
respect to the more recent work of Ishkhanov $et$ $al.$ \cite{Ish:72}, also measure the ${}^{26}$Mg$\left( {\gamma ,pn}
\right)$ cross section. The three experiments agree with each other with respect to the overall width of the
E1--response, and they all display two maxima at 18 MeV and 22 MeV separated by a shallow minimum around 20 MeV. This
is in qualitative agreement with our calculated curves, especially when the THO basis is used and the QRPA spectra is
folded with a Lorentian distribution with $\Gamma$=3 MeV.

Combining the information from Table II and III, one notices that calculations based on a given interaction predict a
rather small shift of the mean energy of the giant dipole resonance, when comparing the two nuclei, $^{24}$Mg and
$^{26}$Mg. This is at variance with experiment, according to which the mean energy shifts down by 1.7 MeV. In this way,
a calculation which is in accordance with the data for $^{24}$Mg, will disagree with data for $^{26}$Mg, and viceversa.

For completeness Fig. \ref{fig Mg24 no soC dipolo} shows the effect of leaving out the spin--orbit and Coulomb part of
the interaction and one sees that this has only a minor effect contrary to the considerable effect obtained for the
quadrupole mode.

\begin{figure}[!!h]
\begin{center}
\includegraphics*[width=0.38\textwidth,angle=0]{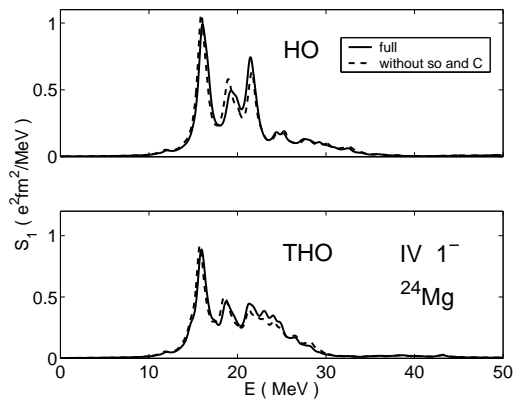}%cut 50
\end{center}
\caption{\footnotesize{The same as Fig. \ref{fig Mg24 no soC} but for IV $1^-$ strength functions..}} \label{fig Mg24
no soC dipolo}
\end{figure}

\subsubsection{IS 0$^+$ modes}

In Fig. \ref{fig Mg24 exp0} we display the experimental fraction of EWSR of Youngblood $et$ $al.$ \cite{Lui.ea:99} and
those obtained in the present work for the IS monopole mode of $^{24}$Mg. The theoretical response shows a two-peak
structure, with a low energy peak well defined around 18 MeV and a fragmented high energy component which appears to be
more sensitive to details and to the choice of the basis. Indeed, in the HO basis there is a significant contribution
around 22 MeV, while in the THO basis this is in the energy range 25--30 MeV. Such peaks are not observed in the
experimental curve, which is quite flat and covers a broad energy interval. Except for the peaks, this spread-out
behavior of the strength function is also found in the calculations. Also for the monopole mode, the calculated EWSR of
90 $\%$ is larger than the experimental value, 72 $\pm$ 10 $\%$. From the similarity of the shape of the curves one can
expect that the calculated centroids and the experimental ones are in good agreement, as given by the numbers of Table
\ref{tab1mg24}.

Fig. \ref{fig Mg26 00} shows the calculated fraction of EWSR for the IS monopole mode of $^{26}$Mg. As for $^{24}$Mg,
the behavior of the response is quite sensitive to the choice of the basis. The curve displayed in the HO basis has a
behavior similar to that obtained by S. P\'{e}ru $et$ $al.$ \cite{Peru.ea:08}, i.e. the resonance has a prominent peak
around 20 MeV and a broad component at higher energy. In the THO basis the low--energy peak of the resonance is shifted
down by 1.5 MeV and there is a significant contribution around 26 MeV.

\begin{figure}[!!h]
\begin{center}
\includegraphics*[width=0.38\textwidth,angle=0]{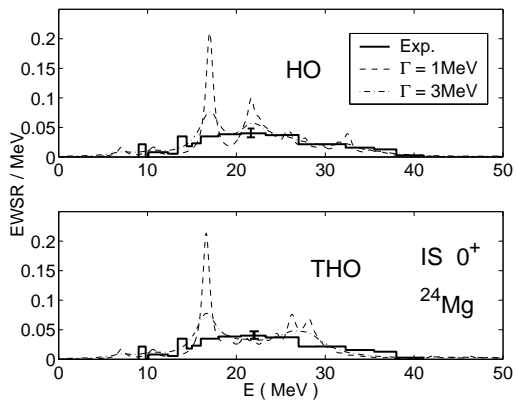}%cut 50
\end{center}
\caption{\footnotesize{The same as Fig. \ref{fig Mg24 exp2} but for fractions of IS monopole EWSR.}} \label{fig Mg24
exp0}
\end{figure}

\begin{figure}[!!h]
\begin{center}
\includegraphics*[width=0.38\textwidth,angle=0]{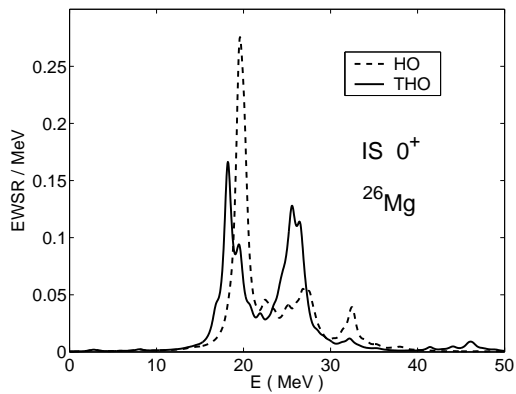}%cut 50
\end{center}
\caption{\footnotesize{Fractions of IS monopole EWSR in ${}^{26}$Mg, in HO basis (dashed curve) and THO (solid curve)
basis.}} \label{fig Mg26 00}
\end{figure}

\subsection{${}^{34}$Mg}

\begin{figure}[!!h]
\begin{center}
\includegraphics*[width=0.38\textwidth,angle=0]{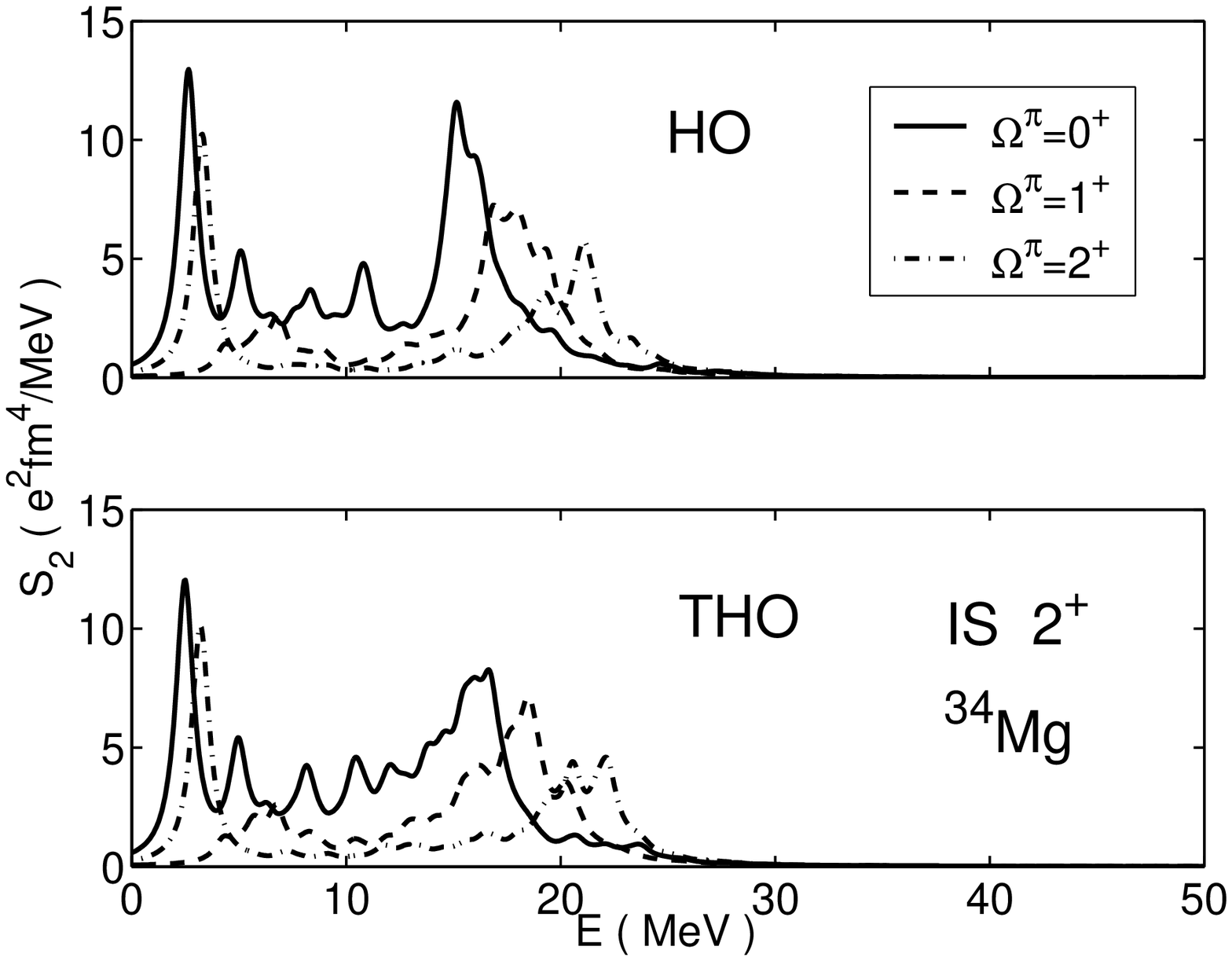}
\end{center}
\caption{\footnotesize{IS $2^+$ strength functions in ${}^{34}$Mg, in the HO (upper panel) and in the THO (lower panel)
basis for the $\Omega^\pi$=$0^+$ (solid curve), $1^+$ (dashed curve) and $2^+$ (dashed-dotted curve) excitations.}}
\label{fig Mg34 2IS}
\end{figure}

Fig. \ref{fig Mg34 2IS} shows the calculated IS quadrupole strength functions for the neutron--rich nucleus ${}^{34}$Mg
in the HO and THO basis. One can see a low--lying peak at 2--3 MeV and a giant resonance at 15--22 MeV. The first
low--lying state belongs to the $\Omega^\pi$=$0^+$ component and it is mainly constructed by the neutron pp excitations
$\left[ {N, n_z , \Lambda , \pm \Omega ^\pi } \right] \equiv \left[ {2,0,2, \pm \frac{3} {2}^ +  } \right]^2$, $\left[
{3,2,1, \pm \frac{3} {2}^ + } \right]^2$, $\left[ {3,3,0, \pm \frac{1} {2}^ - } \right]^2$, the first of them coming
from $\left( 1d_{3/2}\right)^2$ the latter two from $\left( 1f_{7/2}\right)^2$. The features of these transitions are
shown in Tables \ref{tab1 mg34 HO}, \ref{tab1 mg34 THO} for the HO and THO basis respectively. These results are in
fair agreement with those of Table I of ref. \cite{Yosh.Ya:08}. One may notice that the two approaches give basically
the same results and that the difference of about 200 keV between the two energy peaks (2.64 MeV in the HO basis and at
2.48 MeV in the THO basis) is due to the unperturbed 2qp transitions with energy $E_{qpK}+E_{qpK}$. From the values of
these 2qp energy transitions one may conclude that the main contribution of the residual interaction is to shift down
in energy the IS quadrupole response, rather independently on the choice of basis.

\begin{figure}[!!h]
\begin{center}
\includegraphics*[width=0.38\textwidth,angle=0]{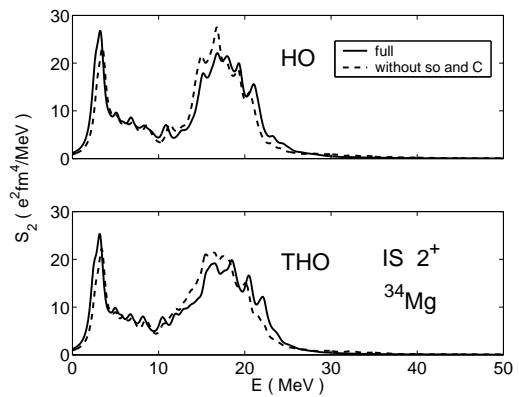}
\end{center}
\caption{\footnotesize{The same as Fig. \ref{fig Mg24 no soC} but for ${}^{34}$Mg.}} \label{fig lorenz2 mg34}
\end{figure}

In Fig. \ref{fig lorenz2 mg34} we show the total isoscalar 2$^+$ response functions in the HO and THO basis. When the
spin--orbit and Coulomb residual interaction are omitted, the low--lying peak belonging to the $\Omega^\pi$=$0^+, 2^+$
components is shifted up in energy by about 0.2 MeV. The giant resonance is instead shifted down in energy by about 0.6
MeV respect to the value obtained in the fully consistent calculation. The behavior of the strength functions
calculated without spin--orbit and Coulomb interactions is quite close to the corresponding calculations by K. Yoshida
$et$ $al.$ \cite{Yosh.ea:08}. The giant resonance of our calculations is instead shifted up in energy by about 1.3 MeV
respect to that in ref. \cite{Yosh.ea:08}. This is coherent with the former comparisons between our ISGQR and those of
K. Yoshida $et$ $al.$ in ${}^{20}$O where a difference of about 0.6 MeV is found, and in ${}^{24}$Mg where there is a
difference of 1.8 MeV. It seems that the blue shift displayed by the peaks in Fig. \ref{fig lorenz2 mg34} as compared
to those obtained in the calculations of K. Yoshida $et$ $al.$ without taking into account spin--orbit and Coulomb
contributions in the residual interaction, increases with the intrinsic deformation of the nuclear system.

%Fig. \ref{fig lorenz2 mg34} shows the total isoscalar 2$^+$ response functions in the HO and THO basis. The behavior is
%quite close to the corresponding calculations by K. Yoshida $et$ $al.$ \cite{Yosh.ea:08}. There is a remarkable
%agreement with respect to the low--lying peak at 2--3 MeV belonging to the $\Omega^\pi$=$0^+, 2^+$ components. The
%giant resonance of our calculations is instead shifted up in energy by about 1.3 MeV respect to that in ref.
%\cite{Yosh.ea:08}. This is coherent with the former comparisons between our ISGQR and those of K. Yoshida $et$ $al.$ in
%${}^{20}$O where a difference of about 0.6 MeV is found, and in ${}^{24}$Mg where there is a difference of 1.8 MeV. It
%seems that the blue shift displayed by the peaks displayed in Fig. \ref{fig lorenz2 mg34} as compared to those obtained
%in the calculations of K. Yoshida $et$ $al.$ calculated without taking into account spin--orbit and Coulomb
%contributions in the residual interaction, increases with the intrinsic deformation of the nuclear system.

\begin{table}
\begin{ruledtabular}
\begin{tabular}{lccc}
%                             &                                                         &                                                          &                                                    \\
%\small{$K$}                  &\small{$\left[ {2,0,2, \pm \frac{3} {2}^ +  } \right]$}  &\small{$\left[ {3,2,1, \pm \frac{3} {2}^ -  } \right]$}   &$\left[ {3,3,0, \pm \frac{1} {2}^ -  } \right]$     \\
\small{$KK$}                   &\small{$\left[ {2,0,2, \pm \frac{3} {2}^ +  } \right]^2$}  &\small{$\left[ {3,2,1, \pm \frac{3} {2}^ -  } \right]^2$}   &$\left[ {3,3,0, \pm \frac{1} {2}^ -  } \right]^2$     \\
%                             &                                                         &                                                          &                                                    \\
%\small{$K'$}                 &\small{$\left[ {2,0,2, \pm \frac{3} {2}^ +  } \right]$}  &\small{$\left[ {3,2,1, \pm \frac{3} {2}^ -  } \right]$}   &$\left[ {3,3,0, \pm \frac{1} {2}^ -  } \right]$     \\
                             &                                                         &                                                          &                                                    \\
\hline
\small{$q_K$}                &$n$                                                      &$n$                                                       &$n$                                                 \\
%\small{$q_{K'}$}             &$n$                                                      &$n$                                                       &$n$                                                 \\
\small{$v_K^2$}              &0.25                                                     &0.68                                                      &0.92                                                \\
%\small{$v^2_{K'}$}           &0.25                                                     &0.68                                                      &0.92                                                \\
\small{$\varepsilon_K$}      &-3.18                                                    &-4.91                                                     &-6.84                                               \\
%\small{$\varepsilon_{K'}$}   &-3.18                                                    &-4.91                                                     &-6.84                                               \\
\small{$X_{KK}$}             &-0.64                                                    &0.65                                                      &0.29                                                \\
\small{$Y_{KK}$}             &-0.08                                                    &0.03                                                      &0.04                                                \\
\small{$E_{qpK}+E_{qpK}$}    &3.79                                                     &3.72                                                      &5.96                                                \\
\end{tabular}
\end{ruledtabular}
\caption{\label{tab1 mg34 HO} QRPA amplitudes $\left[ {N, n_z, \Lambda, \pm \Omega ^\pi } \right]_K^2$ with isospin $q$
($q=n$ for neutrons, $q=p$ for protons) of the IS quadrupole $\Omega ^\pi = 0^+$ mode at 2.64 MeV for the HO basis in
${}^{34}$Mg. The quasi particle occupations $v_K^2$, the single particle energies $\varepsilon_K$ and the 2qp
excitation energies $E_{qpK}+E_{qpK}$, are given. Only components with $X_{KK}^2-Y_{KK}^2 > 0.01$ are listed.}
\end{table}

\begin{table}
\begin{ruledtabular}
\begin{tabular}{lccc}
%                             &                                                         &                                                          &                                                    \\
%\small{$K$}                  &\small{$\left[ {2,0,2, \pm \frac{3} {2}^ +  } \right]$}  &\small{$\left[ {3,2,1, \pm \frac{3} {2}^ -  } \right]$}   &$\left[ {3,3,0, \pm \frac{1} {2}^ -  } \right]$     \\
\small{$KK$}                   &\small{$\left[ {2,0,2, \pm \frac{3} {2}^ +  } \right]^2$}  &\small{$\left[ {3,2,1, \pm \frac{3} {2}^ -  } \right]^2$}   &$\left[ {3,3,0, \pm \frac{1} {2}^ -  } \right]^2$     \\
%                             &                                                         &                                                          &                                                    \\
%\small{$K'$}                 &\small{$\left[ {2,0,2, \pm \frac{3} {2}^ +  } \right]$}  &\small{$\left[ {3,2,1, \pm \frac{3} {2}^ -  } \right]$}   &$\left[ {3,3,0, \pm \frac{1} {2}^ -  } \right]$     \\
                             &                                                         &                                                          &                                                    \\
\hline
\small{$q_K$}                &$n$                                                      &$n$                                                       &$n$                                                 \\
%\small{$q_{K'}$}             &$n$                                                      &$n$                                                       &$n$                                                 \\
\small{$v_K^2$}              &0.24                                                     &0.70                                                      &0.92                                                \\
%\small{$v^2_{K'}$}           &0.25                                                     &0.68                                                      &0.92                                                \\
\small{$\varepsilon_K$}      &-3.20                                                    &-4.91                                                     &-6.84                                               \\
%\small{$\varepsilon_{K'}$}   &-3.20                                                    &-4.91                                                     &-6.84                                               \\
\small{$X_{KK}$}             &-0.64                                                    &0.67                                                      &0.27                                                \\
\small{$Y_{KK}$}             &-0.09                                                    &0.03                                                      &0.04                                                \\
\small{$E_{qpK}+E_{qpK}$}    &3.58                                                     &3.49                                                      &5.72                                                \\
\end{tabular}
\end{ruledtabular}
\caption{\label{tab1 mg34 THO}The same as Table \ref{tab1 mg34 HO} but for the THO basis at 2.48 MeV.}
\end{table}

%%%% CONCLUSIONE %%%%%
\section{Conclusion}\label{concludo}

We described the development and testing of a fully consistent QRPA method to calculate linear response of
axially--symmetric--deformed nuclei employing the canonical HO or THO HFB basis. The same Skyrme force is used in both
HFB and QRPA approaches in all ph, pp and hh channels. The method is applied to study the responses of ${}^{20}$O,
${}^{24}{}^{-}{}^{26}$Mg and ${}^{34}$Mg.

For ${}^{20}$O we showed the effective role of self-consistency in the IS $2^+$ and IV $1^-$ responses. Within this
context, we performed calculations with and without the spin--orbit and Coulomb terms of the residual interaction. One
can conclude that for this spherical nucleus these terms mainly act on the giant resonances, shifting their centroid up
in energy by several hundred keV.

We carried out detailed studies of the IS quadrupole, monopole and IV dipole responses of deformed prolate and oblate
${}^{24}{}^{-}{}^{26}$Mg. A microscopic analysis of the low--lying IS $2^+$ vibrations showed, for the open shell
nucleus ${}^{26}$Mg major contributions to the response function arising from pp transitions, while in the case of the
nucleus ${}^{24}$Mg only ph (RPA) states are found to be important. In the energy region of giant resonances, the role
of deformation manifests itself in the splitting between the different projections which is more pronounced for the
strongly deformed ${}^{24}$Mg (inhomogeneous damping). On top of this splitting comes a fragmentation, especially of
the highest energy projections of the giant modes. These effects lead to broad strength distributions. In this respect,
our calculations confirm earlier theoretical results, and our consistent inclusion of the spin-orbit and Coulomb parts
of the residual interaction only introduces minor changes concerning the giant modes, shifting the centroid by
typically 1 MeV and leaving the widths rather unaffected. The calculated strength functions in regions of giant
vibrations are compared to the available data, yielding generally an overall account of the experimental findings.

For ${}^{34}$Mg, we give a description of the total IS 2$^+$ response concentrating on the microscopic structure of the
low-lying states which are in overall accord with the theoretical results of K. Yoshida $et$ $al.$
\cite{Yosh.ea:08,Yosh.Ya:08}.

We plan to optimize the present code, to be able to perform systematic calculations for both spherical and deformed
systems in particular on the isotopes of Mg and of O. The aim is to extend our analysis to light drip line nuclei, so
as to be able to study also pymgy resonances as well as other collective states typical of these exotic species. From a
technical point of view, we also plan to assess for these exotic nuclear systems the significance of applying the
THO-basis, which should more properly take into account the extended tails of wave functions than the HO basis.

In a general perspective, the present work represents the first, unavoidable step for a consistent and more systematic
study of collective modes in nuclei, in particular light exotic nuclei, taking properly into account also medium
polarization effects. A study of the core polarization effects in Al isotopes has been performed by K. Yoshida
\cite{Yosh.Ya:09h} by employing the quasiparticle-vibration-coupling model on top of the deformed HFB plus QRPA using a
Skyrme interaction.

The plan is to use the resulting states to calculate the role of the exchange of phonons between nucleons moving in
time reversal states close to the Fermi energy has in Cooper pair binding in exotic, deformed (as well as spherical)
nuclei.

%%%%% RINGRAZIAMENTI %%%%%%
\section*{Acknowledgments}

The authors thank J. Dobaczewski for valuable comments and discussions. They also acknowledge J. Terasaki who kindly
gave the great opportunity to apply his QRPA spherical program, M. Stoitsov who provided the last version (101) of the
program HFBTHO, S. P\'{e}ru and K. Yoshida for the interest they have shown for this work. One of the authors (A.P.) is
supported by the Academy of Finland and University of Jyv\"{a}skyl\"{a} within the FIDIPRO programme. C.L. acknowledges
the support by a PhD stipend granted by the Science Faculty of The University of Copenhagen. The numerical calculations
were performed on the supercomputers at the Consorzio Interuniversitario Lombardo Elaborazione Automatica (CILEA),
Segrate, Italy.

%%%%% Appendice %%%%%%
\section*{Appendix: Evaluation of interaction matrix elements}

\def\thesubsection{\Alph{subsection}}
\subsection{Canonical wave-functions}

In this appendix we discuss the procedures used to evaluate some of the characteristic ph effective Skyrme matrix elements for
the canonical wave functions having the general form:

\begin{equation}\label{wf}
\begin{gathered}
  \Phi _K (\textit{\textbf{r}},\boldsymbol\sigma ) \hfill \\
   = \frac{1}
{{\sqrt {2\pi } }}\left[ {\varphi _{K \uparrow } (r_ \bot  ,z)e^{i\Lambda _K^ -  \phi } \left|  \uparrow  \right\rangle  + \varphi _{K \downarrow } (r_ \bot  ,z)e^{i\Lambda _K^ +  \phi } \left|  \downarrow  \right\rangle } \right], \hfill \\
\end{gathered}
\end{equation}

\noindent with spin--up and spin--down components ${\varphi _{K \uparrow } (r_ \bot  ,z)}$, ${\varphi _{K \downarrow }
(r_ \bot ,z)}$ obtained in the present work in the harmonic oscillator basis, i.e. by the associated Laguerre and
Hermite polynomials \cite{Abra:70}. The general form of these matrix elements and their effective Skyrme
parametrization is given by J. Terasaki $et$ $al.$ \cite{Ter.ea:05} in Eqs. (B12) to (B19). Also, for the QRPA
formalism we refer to J. Terasaki $et$ $al.$, Eqs. (A1)--(A6) in ref. \cite{Ter.ea:05}.

Concerning the most simple interactions, namely those proportional to the contact function $\delta (\textit{\textbf{r}} - \textit{\textbf{r}}')$, one has to evaluate the overlaps,

\begin{equation}\label{delta1}
\begin{gathered}
  \left\langle {KK'|\delta (\textit{\textbf{r}} - \textit{\textbf{r}}')|LL'} \right\rangle  \hfill \\
   = \int {dr} \left[ {\Phi _K^* (\textit{\textbf{r}},\boldsymbol\sigma )\Phi _L (\textit{\textbf{r}},\boldsymbol\sigma )} \right]\left[ {\Phi _{K'}^* (\textit{\textbf{r}},\boldsymbol\sigma )\Phi _{L'} (\textit{\textbf{r}},\boldsymbol\sigma )} \right], \hfill \\
\end{gathered}
\end{equation}

\noindent overlaps which in terms of the canonical wave-functions (\ref{wf}) read

\begin{equation}\label{delta2}
\begin{gathered}
  \left\langle {KK'|\delta (\textit{\textbf{r}} - \textit{\textbf{r}}')|LL'} \right\rangle  = \frac{1}
{{2\pi }}\delta _{\Omega _K  + \Omega _{K'} ,\Omega _L  + \Omega _{L'} }  \hfill \\
  \begin{array}{*{20}c}
   {}  \\

 \end{array}  \times \int {r_ \bot  dr_ \bot  dz\left[ {\varphi _{K \uparrow } (r_ \bot  ,z)\varphi _{L \uparrow } (r_ \bot  ,z)} \right.}  \hfill \\
  \begin{array}{*{20}c}
   {}  \\

 \end{array} \left. { + \varphi _{K \downarrow } (r_ \bot  ,z)\varphi _{L \downarrow } (r_ \bot  ,z)} \right] \hfill \\
  \begin{array}{*{20}c}
   {}  \\

 \end{array}  \times \left[ {\varphi _{K' \uparrow } (r_ \bot  ,z)\varphi _{L' \uparrow } (r_ \bot  ,z) + \varphi _{K' \downarrow } (r_ \bot  ,z)\varphi _{L' \downarrow } (r_ \bot  ,z)} \right]. \hfill \\
\end{gathered}
\end{equation}

Next, we turn to the interaction ${\boldsymbol\sigma  \cdot \boldsymbol\sigma '\delta (\textit{\textbf{r}} -
\textit{\textbf{r}}')}$. Here, the spin product is written in terms of lowering and raising spin operators $\sigma _ +$ and
$\sigma _{ - }$

\begin{equation}\label{sigma iddotprod}
\boldsymbol\sigma  \cdot \boldsymbol\sigma' = \frac{1} {2}\left[ {\sigma_+  \sigma_-' + \sigma_-  \sigma_+'} \right] + \sigma_z \sigma_z'.
\end{equation}

\noindent The algebra containing the $\delta (\textit{\textbf{r}} - \textit{\textbf{r}}')$ function is the same as
above, leading to the result

\begin{equation}\label{delta sigma}
\begin{gathered}
  \left\langle {KK'|\boldsymbol\sigma  \cdot \boldsymbol\sigma '\delta (\textit{\textbf{r}} -
\textit{\textbf{r}}')|LL'} \right\rangle  = \frac{1}
{{2\pi }}\delta _{\Omega _K  + \Omega _{K'} ,\Omega _L  + \Omega _{L'} }  \hfill \\
  \begin{array}{*{20}c}
   {}  \\

 \end{array}  \times \int {r_ \bot  dr_ \bot  dz\left\{ {\left[ {\varphi _{K \uparrow } (r_ \bot  ,z)\varphi _{L \uparrow } (r_ \bot  ,z)} \right.} \right.}  \hfill \\
  \begin{array}{*{20}c}
   {}  \\

 \end{array}  \left. { - \varphi _{K \downarrow } (r_ \bot  ,z)\varphi _{L \downarrow } (r_ \bot  ,z)} \right] \hfill \\
  \begin{array}{*{20}c}
   {}  \\

 \end{array}  \times \left[ {\varphi _{K' \uparrow } (r_ \bot  ,z)\varphi _{L' \uparrow } (r_ \bot  ,z) - \varphi _{K' \downarrow } (r_ \bot  ,z)\varphi _{L' \downarrow } (r_ \bot  ,z)} \right] \hfill \\
  \begin{array}{*{20}c}
   {}  \\

 \end{array}  + 2\left[ {\varphi _{K \uparrow } (r_ \bot  ,z)\varphi _{K' \downarrow } (r_ \bot  ,z)\varphi _{L \downarrow } (r_ \bot  ,z)\varphi _{L' \uparrow } (r_ \bot  ,z)} \right. \hfill \\
  \left. {\left. {\begin{array}{*{20}c}
   {}  \\

 \end{array}  + \varphi _{K \downarrow } (r_ \bot  ,z)\varphi _{K' \uparrow } (r_ \bot  ,z)\varphi _{L \uparrow } (r_ \bot  ,z)\varphi _{L' \downarrow } (r_ \bot  ,z)} \right]} \right\}, \hfill \\
\end{gathered}
\end{equation}

\noindent where one can recognize the action of $\sigma _z \sigma _z '$ in the first term, and likewise $\frac{1}
{2}{\sigma_+  \sigma_-'}$ and $\frac{1} {2}{\sigma_-  \sigma_+'}$ in the two subsequent terms.

It turns out that the momentum dependent parts of the residual interaction, involving differentiation in terms of the
$\textit{\textbf{k}}$ and $\textit{\textbf{k}}^{\dag}$ operators, give rise to quite many terms. Expressing one of
these in full length

\begin{equation}\label{operatore k2}
\textit{\textbf{k}}^2  =  - \frac{1} {4}\nabla^2  - \frac{1} {4}\nabla'^2  + \frac{1} {2}\nabla  \cdot \nabla',
\end{equation}

\noindent one obtains

\begin{widetext}
\begin{equation}\label{delta k2 tutto}
\begin{gathered}
  \left\langle {KK'|\delta (\textit{\textbf{r}} -
\textit{\textbf{r}}')\textit{\textbf{k}}^2 |LL'} \right\rangle  \hfill \\
   =  - \frac{1}
{{8\pi }}\delta _{\Omega _K  + \Omega _{K'} ,\Omega _L  + \Omega _{L'} } \int {r_ \bot  dr_ \bot  dz\left[ {\varphi _{K
\uparrow } (r_ \bot  ,z)\left( {\frac{1} {{r_ \bot  }}\frac{\partial } {{\partial r_ \bot  }} + \frac{{\partial ^2 }}
{{\partial r_ \bot ^2 }} - \frac{{\left( {\Omega _L  - \frac{1} {2}} \right)^2 }} {{r_ \bot ^2 }} + \frac{{\partial ^2
}}
{{\partial z^2 }}} \right)} \right.} \varphi _{L \uparrow } (r_ \bot  ,z) \hfill \\
  \begin{array}{*{20}c}
   {}  \\

 \end{array}  + \left. {\varphi _{K \downarrow } (r_ \bot  ,z)\left( {\frac{1}
{{r_ \bot  }}\frac{\partial } {{\partial r_ \bot  }} + \frac{{\partial ^2 }} {{\partial r_ \bot ^2 }} - \frac{{\left(
{\Omega _L  + \frac{1} {2}} \right)^2 }} {{r_ \bot ^2 }} + \frac{{\partial ^2 }}
{{\partial z^2 }}} \right)\varphi _{L \downarrow } (r_ \bot  ,z)} \right] \hfill \\
  \begin{array}{*{20}c}
   {}  \\

 \end{array}  \times \left[ {\varphi _{K' \uparrow } (r_ \bot  ,z)\varphi _{L' \uparrow } (r_ \bot  ,z) + \varphi _{K' \downarrow } (r_ \bot  ,z)\varphi _{L' \downarrow } (r_ \bot  ,z)} \right] \hfill \\
  \begin{array}{*{20}c}
   {}  \\

 \end{array}  - \frac{1}
{{8\pi }}\delta _{\Omega _K  + \Omega _{K'} ,\Omega _L  + \Omega _{L'} } \int {r_ \bot  dr_ \bot  dz\left[ {\varphi _{K \uparrow } (r_ \bot  ,z)\varphi _{L \uparrow } (r_ \bot  ,z) + \varphi _{K \downarrow } (r_ \bot  ,z)\varphi _{L \downarrow } (r_ \bot  ,z)} \right]}  \hfill \\
  \begin{array}{*{20}c}
   {}  \\

 \end{array}  \times \left[ {\varphi _{K' \uparrow } (r_ \bot  ,z)\left( {\frac{1}
{{r_ \bot  }}\frac{\partial } {{\partial r_ \bot  }} + \frac{{\partial ^2 }} {{\partial r_ \bot ^2 }} - \frac{{\left(
{\Omega _{L'}  - \frac{1} {2}} \right)^2 }} {{r_ \bot ^2 }} + \frac{{\partial ^2 }}
{{\partial z^2 }}} \right)} \right.\varphi _{L' \uparrow } (r_ \bot  ,z) \hfill \\
  \begin{array}{*{20}c}
   {}  \\

 \end{array}  + \left. {\varphi _{K' \downarrow } (r_ \bot  ,z)\left( {\frac{1}
{{r_ \bot  }}\frac{\partial } {{\partial r_ \bot  }} + \frac{{\partial ^2 }} {{\partial r_ \bot ^2 }} - \frac{{\left(
{\Omega _{L'}  + \frac{1} {2}} \right)^2 }} {{r_ \bot ^2 }} + \frac{{\partial ^2 }}
{{\partial z^2 }}} \right)\varphi _{L' \downarrow } (r_ \bot  ,z)} \right] \hfill \\
  \begin{array}{*{20}c}
   {}  \\

 \end{array}  + \frac{1}
{{4\pi }}\delta _{\Omega _K  + \Omega _{K'} ,\Omega _L  + \Omega _{L'} } \int {r_ \bot  dr_ \bot  dz} \left\{ {\varphi
_{K \uparrow } (r_ \bot  ,z)\varphi _{K' \uparrow } (r_ \bot  ,z)\left[ {\frac{{\partial \varphi _{L \uparrow } (r_
\bot  ,z)}} {{\partial r_ \bot  }}\frac{{\partial \varphi _{L' \uparrow } (r_ \bot  ,z)}}
{{\partial r_ \bot  }}} \right.} \right. \hfill \\
  \left. {\begin{array}{*{20}c}
   {\begin{array}{*{20}c}
   {}  \\

 \end{array}  - \frac{{\left( {\Omega _L  - \frac{1}
{2}} \right)\left( {\Omega _{L'}  - \frac{1} {2}} \right)}} {{r_ \bot ^2 }}\varphi _{L \uparrow } (r_ \bot  ,z)\varphi
_{L' \uparrow } (r_ \bot  ,z) + \frac{{\partial \varphi _{L \uparrow } (r_ \bot  ,z)}} {{\partial z}}\frac{{\partial
\varphi _{L' \uparrow } (r_ \bot  ,z)}}
{{\partial z}}}  \\

 \end{array} } \right] \hfill \\
  \begin{array}{*{20}c}
   {}  \\

 \end{array}  + \varphi _{K \uparrow } (r_ \bot  ,z)\varphi _{K' \downarrow } (r_ \bot  ,z)\left[ {\frac{{\partial \varphi _{L \uparrow } (r_ \bot  ,z)}}
{{\partial r_ \bot  }}\frac{{\partial \varphi _{L' \downarrow } (r_ \bot  ,z)}} {{\partial r_ \bot  }} - \frac{{\left(
{\Omega _L  - \frac{1} {2}} \right)\left( {\Omega _{L'}  + \frac{1} {2}} \right)}}
{{r_ \bot ^2 }}\varphi _{L \uparrow } (r_ \bot  ,z)\varphi _{L' \downarrow } (r_ \bot  ,z)} \right. \hfill \\
  \begin{array}{*{20}c}
   {}  \\

 \end{array}  + \left. {\frac{{\partial \varphi _{L \uparrow } (r_ \bot  ,z)}}
{{\partial z}}\frac{{\partial \varphi _{L' \downarrow } (r_ \bot  ,z)}} {{\partial z}}} \right] + \varphi _{K
\downarrow } (r_ \bot  ,z)\varphi _{K' \uparrow } (r_ \bot  ,z)\left[ {\frac{{\partial \varphi _{L \downarrow } (r_
\bot  ,z)}} {{\partial r_ \bot  }}\frac{{\partial \varphi _{L' \uparrow } (r_ \bot  ,z)}}
{{\partial r_ \bot  }}} \right. \hfill \\
  \left. {\begin{array}{*{20}c}
   {}  \\

 \end{array}  - \frac{{\left( {\Omega _L  + \frac{1}
{2}} \right)\left( {\Omega _{L'}  - \frac{1} {2}} \right)}} {{r_ \bot ^2 }}\varphi _{L \downarrow } (r_ \bot ,z)\varphi
_{L' \uparrow } (r_ \bot  ,z) + \frac{{\partial \varphi _{L \downarrow } (r_ \bot  ,z)}} {{\partial z}}\frac{{\partial
\varphi _{L' \uparrow } (r_ \bot  ,z)}}
{{\partial z}}} \right] \hfill \\
  \begin{array}{*{20}c}
   {}  \\

 \end{array}  + \varphi _{K \downarrow } (r_ \bot  ,z)\varphi _{K' \downarrow } (r_ \bot  ,z)\left[ {\frac{{\partial \varphi _{L \downarrow } (r_ \bot  ,z)}}
{{\partial r_ \bot  }}\frac{{\partial \varphi _{L' \downarrow } (r_ \bot  ,z)}} {{\partial r_ \bot  }} - \frac{{\left(
{\Omega _L  + \frac{1} {2}} \right)\left( {\Omega _{L'}  + \frac{1} {2}} \right)}}
{{r_ \bot ^2 }}\varphi _{L \downarrow } (r_ \bot  ,z)\varphi _{L' \downarrow } (r_ \bot  ,z)} \right. \hfill \\
  \left. {\left. {\begin{array}{*{20}c}
   {}  \\

 \end{array}  + \frac{{\partial \varphi _{L \downarrow } (r_ \bot  ,z)}}
{{\partial z}}\frac{{\partial \varphi _{L' \downarrow } (r_ \bot  ,z)}}
{{\partial z}}} \right]} \right\}. \hfill \\
\end{gathered}
\end{equation}
\end{widetext}

\def\thesubsection{\Alph{subsection}}
\subsection{Spin-orbit interaction}

The spin-orbit term ${i(\boldsymbol\sigma  + \boldsymbol\sigma ') \cdot \textit{\textbf{k}}^\dag \times \delta
(\textit{\textbf{r}} - \textit{\textbf{r}}')\textit{\textbf{k}}}$ of the residual interaction is likely the most
involved to evaluate. However, it becomes conceptually simple when it is interpreted as a volume product of three
vectors $\textit{\textbf{A}} \cdot \left( {\textit{\textbf{B}} \times \textit{\textbf{C}}} \right)$. Next, the
cartesian components of the vectors are replaced by the components of the spherical tensors of rank 1 \cite{Boh:75a}.
This bears some resemblance to the treatment in the spherical case \cite{Ter.ea:05}. In the deformed case, the spin
spherical tensors raising and lowering operators, and the spherical tensors of the differential
operators also acquire a more simple form than their cartesian counterparts, as they can be expressed in terms of raising
and lowering operators as

\begin{equation}\label{nabla piu}
\nabla _ +   = e^{i\phi } \left( {\frac{\partial } {{\partial r_ \bot  }} + \frac{i} {{r_ \bot  }}\frac{\partial }
{{\partial \phi }}} \right),
\end{equation}

and

\begin{equation}\label{nabla meno}
\nabla _ -   = e^{ - i\phi } \left( {\frac{\partial } {{\partial r_ \bot  }} - \frac{i} {{r_ \bot  }}\frac{\partial }
{{\partial \phi }}} \right).
\end{equation}

\noindent Still, the volume product contains six terms, namely

\begin{widetext}

\begin{equation}\label{matrice so}
\begin{gathered}
  \left\langle {KK'|i(\boldsymbol\sigma  + \boldsymbol\sigma ') \cdot \textit{\textbf{k}}^\dag   \times \delta (\textit{\textbf{r}} - \textit{\textbf{r}}')\textit{\textbf{k}}|LL'} \right\rangle  = \frac{1}
{8}\left[ {\left\langle {KK'} \right|\left( {\sigma _ +   + \sigma _ +  '} \right)\left( {\nabla _z  - \nabla _z '} \right)\delta (\textit{\textbf{r}} - \textit{\textbf{r}}')\left( {\nabla _ -   - \nabla _ -  '} \right)\left| {LL'} \right\rangle } \right. \hfill \\
   - \left\langle {KK'} \right|\left( {\sigma _ +   + \sigma _ +  '} \right)\left( {\nabla _ -   - \nabla _ -  '} \right)\delta (\textit{\textbf{r}} - \textit{\textbf{r}}')\left( {\nabla _z  - \nabla _z '} \right)\left| {LL'} \right\rangle  + \left\langle {KK'} \right|\left( {\sigma _z  + \sigma _z '} \right)\left( {\nabla _ -   - \nabla _ -  '} \right)\delta (\textit{\textbf{r}} - \textit{\textbf{r}}')\left( {\nabla _ +   - \nabla _ +  '} \right)\left| {LL'} \right\rangle  \hfill \\
   - \left\langle {KK'} \right|\left( {\sigma _z  + \sigma _z '} \right)\left( {\nabla _ +   - \nabla _ +  '} \right)\delta (\textit{\textbf{r}} - \textit{\textbf{r}}')\left( {\nabla _ -   - \nabla _ -  '} \right)\left| {LL'} \right\rangle  + \left\langle {KK'} \right|\left( {\sigma _ -   + \sigma _ -  '} \right)\left( {\nabla _ +   - \nabla _ +  '} \right)\delta (\textit{\textbf{r}} - \textit{\textbf{r}}')\left( {\nabla _z  - \nabla _z '} \right)\left| {LL'} \right\rangle  \hfill \\
  \left. { - \left\langle {KK'} \right|\left( {\sigma _ -   + \sigma _ -  '} \right)\left( {\nabla _z  - \nabla _z '} \right)\delta (\textit{\textbf{r}} - \textit{\textbf{r}}')\left( {\nabla _ +   - \nabla _ +  '} \right)\left| {LL'} \right\rangle } \right]. \hfill \\
\end{gathered}
\end{equation}

\end{widetext}

\noindent In the above relation, the operators to the left of $\delta (\textit{\textbf{r}} -
\textit{\textbf{r}}')$ act on $K$, $K'$, while the operators to the right of $\delta (\textit{\textbf{r}} -
\textit{\textbf{r}}')$ act on $L$, $L'$. The sum over the six terms of Eq. (\ref{matrice so}) corresponds to a sum over
the six permutations of the set $\left\{ { + ,z, - } \right\}$.

\subsection{Coulomb interaction}

Due to the short range character of the nuclear residual interaction, which is expressed through the $\delta
(\textit{\textbf{r}} - \textit{\textbf{r}}')$ functions, the four-dimensional integrals $\int {r_\bot  dr_\bot dzr_\bot' dr_\bot' dz'}$ are directly replaced by two-dimensional integrals. However, for the direct term of the Coulomb
interaction $V_{direct}^{eff}  = \frac{{e^2 }} {{\left| {\textit{\textbf{r}} - \textit{\textbf{r}}'} \right|}}$ one needs to carry out the full integration. In order to exploit the cylindrical symmetry,
we make use of an expansion method applied to astrophysical problems by Cohl and Tohline
\cite{Cohl:99,Cohl:01}

\begin{equation}\label{cylindrical expansion}
\frac{1} {{\left| {\textit{\textbf{r}} - \textit{\textbf{r}}'} \right|}} = \frac{1} {{\pi \sqrt {r_ \bot  r_ \bot  '}
}}\sum\limits_{m =  - \infty }^\infty {Q_{m - \frac{1} {2}} \left( \chi_-  \right)e^{im\left( {\phi  - \phi '} \right)}
} ,
\end{equation}

\noindent where

\begin{equation}\label{chi}
\chi_-  \equiv \frac{{r_ \bot ^2  + {r'}_ \bot ^2  + \left( {z - z'} \right)^2 }} {{2r_ \bot  r_ \bot  '}},
\end{equation}

\noindent Here ${Q_{m - \frac{1} {2}} }$ is a Legendre function of the second kind of half-integer degree
\cite{Abra:70}. We have checked the convergence of this expansion, which is quite rapid when the parameter $\chi$ is
not too close to 1. Inserting the canonical wave-functions, the integrals over the azimuthal angle select the order $m$
of the projection of the angular momentum on the $z$-axis, such that $m$ = $\Omega _{K} - \Omega _{L} = \Omega _{L'} -
\Omega _{K'}$. The functions ${Q_{m - \frac{1} {2}} }$ can readily be evaluated by simple integrals and tabulated over
a suitable range. In the calculation of the matrix element

\begin{equation}\label{coulomb1}
\begin{gathered}
  \left\langle {KK'} \right|V_{direct}^{eff} \left| {LL'} \right\rangle  = \iint {d\textit{\textbf{r}}d\textit{\textbf{r}}'} \hfill \\
  \begin{array}{*{20}c}
   {}  \\

 \end{array}  \times \Phi _K^* (\textit{\textbf{r}},\boldsymbol\sigma )\Phi _{K'}^* (\textit{\textbf{r}}',\boldsymbol\sigma )\frac{{e^2 }}
{{\left| {\textit{\textbf{r}} - \textit{\textbf{r}}'} \right|}}\Phi _L (\textit{\textbf{r}},\boldsymbol\sigma )\Phi _{L'} (\textit{\textbf{r}}',\boldsymbol\sigma ), \hfill \\
\end{gathered}
\end{equation}

\noindent it is important to take into account the symmetry of the wave-functions

\begin{equation}\label{simmetria}
\Phi _K ( - \textit{\textbf{r}},\boldsymbol\sigma ) = \pi _K \Phi _K (\textit{\textbf{r}},\boldsymbol\sigma ),
\end{equation}

\noindent $\pi _K  =  \pm 1$ being the parity of the state $K$ depending on the sign of $z$. By exploiting these symmetries one needs only to carry out integration over $z, z' > 0$, aside from integration over the
two angles $\phi$ and $\phi'$

\begin{equation}\label{coulomb fine}
\begin{gathered}
  \left\langle {KK'} \right|V_{direct}^{eff} \left| {LL'} \right\rangle  = 2e^2 \left( {2\pi } \right)^2 \delta _{m,\Omega _K  - \Omega _L }  \cdot \delta _{m,\Omega _{L'}  - \Omega _{K'} }  \hfill \\
  \begin{array}{*{20}c}
   {}  \\

 \end{array}  \times \int\limits_0^\infty  {r_ \bot  dr_ \bot  } \int\limits_0^\infty  {r_ \bot  'dr_ \bot  '} \int\limits_0^\infty  {dz} \int\limits_0^\infty  {dz'}  \hfill \\
  \begin{array}{*{20}c}
   {}  \\

 \end{array}  \times \left[ {\varphi _{K \uparrow } (r_ \bot  ,z)\varphi _{L \uparrow } (r_ \bot  ,z) + \varphi _{K \downarrow } (r_ \bot  ,z)\varphi _{L \downarrow } (r_ \bot  ,z)} \right] \hfill \\
  \begin{array}{*{20}c}
   {}  \\

 \end{array}  \times \frac{{Q_{m - \frac{1}
{2}} \left( {\chi _ -  } \right) + \pi _{K'}  \cdot \pi _{L'}  \cdot \left( { - 1} \right)^m Q_{m - \frac{1} {2}}
\left( {\chi _ +  } \right)}}
{{\pi \sqrt {r_ \bot  r_ \bot  '} }} \hfill \\
  \begin{array}{*{20}c}
   {}  \\

 \end{array}  \times \left[ {\varphi _{K' \uparrow } (r_ \bot  ',z')\varphi _{L' \uparrow } (r_ \bot  ',z') + \varphi _{K' \downarrow } (r_ \bot  ',z')\varphi _{L' \downarrow } (r_ \bot  ',z')} \right], \hfill \\
\end{gathered}
\end{equation}

\noindent with

\begin{equation}\label{chi piu}
\chi _ +  = \frac{{r_ \bot ^2  + {r'}_ \bot ^2  + \left( {z + z'} \right)^2 }} {{2r_ \bot  r_ \bot '}}.
\end{equation}

\noindent Since $\pi _{K'}  \cdot \pi _{L'}  \cdot \left( { - 1} \right)^m$ does not depend on the projections ${\Omega
_K  - \Omega _L }$, ${\Omega _{L'} - \Omega _{K'} }$, this factor is the same for spin--up and spin--down wave
functions.

\clearpage
\bibliographystyle{apsrev}

\end{document}